\documentclass[onecolumn]{IEEEtran}
\usepackage[dvips]{graphicx}
\usepackage{amssymb}

\def\bfx{{\bf x}}

\def\bfw{{\bf w}}
\def\bfone{{\bf 1}}
\newcommand{\sizex}[1]{\|#1\|}
\newcommand{\lengthx}[1]{\vert #1 \vert}
\newcommand{\intersect}[2]{#1 \cap #2 \neq \emptyset}
\newcommand{\arc}[2]{l_{\partial #1 \cap #2 }}
\newcommand{\eqFirst}[3]{\sizex{#1\cap #3}+\hat r(#2)\arc{#1}{#3}}
\newcommand{\eqSecond}[3]{\sizex{#2\cap #3}-\hat r(#1)\arc{#2}{#3}}

\newtheorem{result}{\bf{Result}}

\def\proof{\noindent {\bf Proof:~}}
\def\defeq{\stackrel{\mathrm{def}}{=}}

\newcommand{\OMIT}[1]{}

\begin{document}
%

\title{Theoretical Evaluation of Offloading through Wireless LANs}
%
%
%
%
%

%
\author{
%
%
Hiroshi Saito and Ryoichi Kawahara\\
NTT Network Technology Laboratories\\
3-9-11, Midori-cho, Musashino-shi,
Tokyo 180-8585, Japan. \\
URL: http://www9.plala.or.jp/hslab/\\
Phone: +81 422 59 4300, Fax: +81 422 59 6364.\\
saito.hiroshi@lab.ntt.co.jp, kawahara.ryoichi@lab.ntt.co.jp
}

\maketitle
\begin{abstract}
Offloading of cellular traffic through a wireless local area network (WLAN) is theoretically evaluated.
First, empirical data sets of the locations of WLAN internet access points are analyzed and
an inhomogeneous Poisson process consisting of high, normal, and low density regions is proposed as a spatial point process model for these configurations.
Second, performance metrics, such as mean available bandwidth for a user and the number of vertical handovers, are evaluated for the proposed model through geometric analysis.
Explicit formulas are derived for the metrics, although they depend on many parameters such as the number of WLAN access points, the shape of each WLAN coverage region, the location of each WLAN access point, the available bandwidth (bps) of the WLAN, and the shape and available bandwidth (bps) of each subregion identified by the channel quality indicator in a cell of the cellular network.
Explicit formulas strongly suggest that the bandwidth a user experiences does not depend on the user mobility.
This is because the bandwidth available by a user who does not move and that available by a user who moves are the same or approximately the same as a probabilistic distribution.
Numerical examples show that parameters, such as the size of regions where placement of WLAN access points is not allowed and the mean density of WLANs in high density regions, have a large impact on performance metrics.
In particular, a homogeneous Poisson process model as the WLAN access point location model largely overestimates the mean available bandwidth for a user and the number of vertical handovers.
The overestimated mean available bandwidth is, for example, about 50\% in a certain condition.
\end{abstract}




\section{Introduction}
Due to the surge in data traffic, cellular network operators need to make large investments in their networks.
For example, AT\&T acknowledged a 50-fold growth in wireless data traffic in a 3-year period, and KT, the largest network operator in Korea, experienced a 10-fold data traffic increase in its wideband code-division multiple access (WCDMA) network \cite{commag}.
Although these operators invested extensively in their cellular networks, further efforts are necessary. The provision of wireless local area network (WLAN) access points (APs) is expected to be one of the most promising ways of mitigating the surge in traffic in cellular networks \cite{offload},\cite{offloadTutorial}.
In addition to the deployment of WLAN APs by cellular network operators, independent operators also provide these points in order to offer Internet access services and obtain subscriber fees from users.
Generally, WLAN internet access services are cheaper and faster than Internet access through cellular networks, although their coverage regions are limited.
Thus, vertical handover (handoff) can occur at the boundary of the WLAN coverage region between the WLAN and cellular network.

Although the relationship between cellular networks and public WLAN services is important, quantitative analysis of this relationship is, as far as we know, very limited.
Choi et al. \cite{commag} compared the traffic growth of KT's WCDMA, WiMAX, and WLAN networks in 2010 and included some quantitative information.
However, they describe the situation only from a macroscopic point of view and do not include any information for each WLAN AP (microscopic point of view) or any theoretical work.

This paper examines a theoretical evaluation of the offloading of cellular networks through public WLAN services and consists of two parts.
The first part uses empirical data on the locations of public WLAN internet APs and analyzes them as a spatial point process.
An inhomogeneous Poisson process is proposed as a model of the configurations of public WLAN Internet APs.
Based on the proposed model, the second part analyzes the performance metrics for the inhomogeneous Poisson process model through integral geometry and derives the formulas for the performance metrics.

Much progress has been made in spatial characterization techniques over the past few decades, and these techniques have been applied in many fields including epidemiological analysis, earthquake occurrence analysis, natural resource distribution, geological analysis, agricultural production, and biological analysis \cite{book2}, \cite{book3}, \cite{book_j}, \cite{spatstat}. However, research related to the first part, that is, spatial characterization of networks including APs and base station placement, has not been sufficiently investigated.
Riihij\"arvi et al. investigated spatial characterization of wireless systems \cite{rii2007}, \cite{rii2010}, \cite{population}.
They analyzed the spatial structure of WLAN AP locations on the east and west coasts of the USA \cite{rii2007}. They found that measured AP locations feature power-law or scale-free behavior in their correlation structures.
They analyzed WLAN AP location data and insisted that the Geyer saturation model as a spatial point process fits the actual data \cite{rii2010}. 
Michalopoulou et al. \cite{population} quantified the dependence between the node distributions of wireless networks (second and third generation (2G/3G) cellular networks) and the underlying population densities.
They showed significant statistical similarities between the locations of 2G base stations and population since the deployment of 2G base stations is complete.
In addition, Andrews et al. \cite{baccelli11} compared the coverage based on the placement of actual base stations and those derived by the Poisson point process.
However, the main focus of this study is the derivation of the signal-to-interference-and-noise ratio and information on the actual base station locations or modeling for them is lacking.
Even when we consider wired access networks, we can see that spatial characterization models have not been extensively studied.
Gloaguen et al. \cite{itc}, \cite{access} focused on the fact that the actual configuration of the wired subscriber network strongly depends on the physical route of roads and derives its stochastic model based on the road configuration model.
Their results enable us to remove time-consuming tasks, such as inputting road data or other geographical information, when we conduct simulations for network evaluation.

Integral geometry (geometric probability) used in the second part of this paper is a mathematical method for evaluating the measures in which a certain set (normally a subset of a plane) satisfies certain characteristics and has been included in several papers regarding network related issues.
For example, a series of papers \cite{infocom}, \cite{mobileComp}, \cite{signal} based on analysis using integral geometry proposed shape estimation methods for a target object based on reports from sensor nodes whose locations are unknown.
Lazos et al. \cite{detection} and Lazos and Poovendran \cite{lazos} directly applied the results of the integral geometry discussed in Chapter 5 Section 6.7 of \cite{Santalo} in an analysis of detecting an object moving in a straight line and in evaluating the probability of $k$-coverage. Kwon and Shroff \cite{routing} also applied integral geometry in an analysis of straight line routing, which is an approximation of shortest path routing, and Choi and Das \cite{energy} used it to select sensors in energy-conserving data gathering.
Currently, integral geometry is also being applied to network survivability studies \cite{failureINFOCOM}, \cite{disaster}.

This paper is organized as follows.
In Section 2, empirical data sets of locations of public WLAN Internet APs are analyzed.
Section 3 provides basic formulas used in the analysis in the following sections.
A model proposed based on the results of Section 2 is described in Section 4.
Sections 5 and 6 evaluate performance metrics before and after introducing WLANs, respectively.
Numerical examples are given in Section 7, and a conclusion is given in Section 8.

\section{Empirical data analysis}
\begin{figure}[tb] 
\begin{center} 
\includegraphics[width=8.5cm,clip]{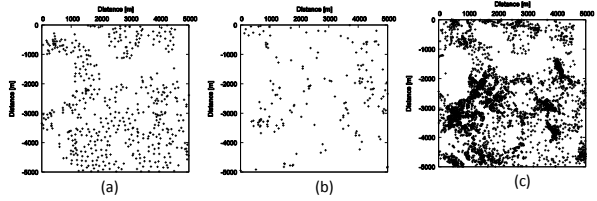} 
\caption{Locations of public WLAN access points} 
\label{map} 
\end{center} 
\end{figure}

We used the location data of public WLAN APs of three operators (a), (b), (c) in Tokyo (Fig. \ref{map}).
(These data were obtained on April 20th, 2012 \cite{livedoor}, October 7th, 2011 \cite{ntt}, and October 20th, 2011 \cite{softB}.)
Figure \ref{map} shows three graphs, which respectively correspond to the three individual operators.
The upper left corners of the graphs represent Shinjuku, one of the busiest regions in Tokyo. We used the location data in this 5 $\times$ 5 km square region.

In the remainder of this section, we investigate the spatial point process model for the AP locations. First, we investigate the hypothesis that they follow a homogeneous Poisson process. Because the statistics using empirical data reject the hypothesis and show that they are more clustered, we also investigate a model that is more clustered than a homogeneous Poisson process.  
Although there are many possible spatial point processes, we propose to adopt an inhomogeneous Poisson process for the model because the number of APs deployed by individual operators at each subregion show high cross-correlations.

\subsection{Test of homogeneous Poisson process}
\begin{figure}[tb] 
\begin{center} 
\includegraphics[width=6cm,clip]{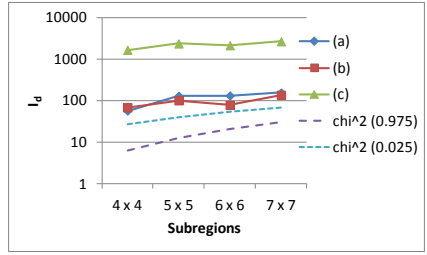} 
\caption{Test using $I_d$.}
\label{I_d} 
\end{center} 
\end{figure}

We conducted a significance test of the null hypothesis of a homogeneous Poisson process.
We divided the entire region into subregions and counted the number of APs in each region.
Let $a_i(j)$ be the number of APs provided by the $j$-th operator in the $i$-th subregion, $A_j$ be the sample mean of the number of APs provided by the $j$-th operator ($A_j\defeq\sum_{i=1}^{n_a}a_i(j)/n_a$), and $V_j$ be its sample variance($V_j\defeq\sum_{i=1}^{n_a}(a_i(j)-A_j))^2/(n_a-1)$), where $n_a$ is the number of subregions.
The index of dispersion $I_d(j) $ proposed by Sachs was applied to the WLAN APs deployed by the $j$-th operator: $I_d(j)\defeq (n_a-1)V_j/A_j$ (p.54 in \cite{book_j}). For a significance test of the null hypothesis of a homogeneous Poisson process, $I_d(j)$ follows a $\chi^2$ distribution with the $n_a-1$ degree of freedom (\cite{book_j}, p.104 in \cite{book3}).
The index $I_d$ for each operator and two-side 5\% bounds of $\chi^2$ for the hypothesis of a homogeneous Poisson process are plotted in Fig. \ref{I_d}.
The results in Fig. \ref{I_d} indicate that we can reject the hypothesis of a homogeneous Poisson process of significance level 0.05. 
That is, $\{\bfw_i\}_i$ is more clustered than a homogeneous Poisson process, where $\bfw_i$ is the location of the $i$-th WLAN AP.

\subsection{Inhomogeneous Poisson process}
\begin{figure}[tb] 
\begin{center} 
\includegraphics[width=6cm,clip]{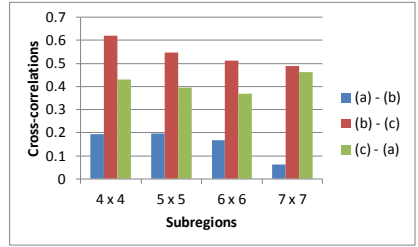} 
\caption{Cross correlations of $\{a_i(j)\}_i$ and $\{a_i(k)\}_i$} \label{cross} 
\end{center} 
\end{figure}

By carefully observing the maps in Fig. \ref{map}, we can see that all three operators have some common sparse regions.
These are regions such as parks or shrines; therefore,
WLAN APs are not allowed or it is practically impossible to place them in these regions.
We should adopt a model that can describe this fact.
Similarly, the operators also have busy regions in common, for example, regions around train/subway stations.
To evaluate whether there are common sparse/busy regions, the
cross-correlations $c(j,k)$ of $\{a_i(j)\}_i$ and $\{a_i(k)\}_i$ were
evaluated, where
$c(j,k)$ is defined as follows: $c(j,k)\defeq \frac{\sum_i (a_i(j)-A_j)(a_i(k)-A_k)}{(n_a-1)\sqrt{V_jV_k}}$.
Figure \ref{cross} shows that there are non-negligible cross correlations, particularly between (b) and (c) and between (c) and (a).

Spatial point process models can be classified into two types.
In the first type, the placement of WLAN APs is assumed to be mainly
determined by the locations of other WLAN APs.
We call these internal models.
Typical examples of internal models are Matern, simple sequential inhibition, and Gibbs point processes \cite{book2},\cite{book_j}.
In the other type, which we call external models, the placement of 
APs is assumed to be mainly determined by the
location features of the APs themselves, rather than the locations of other APs.
An inhomogeneous Poisson process,
i.e., a Poisson process with inhomogeneous intensity, is a typical example of an external model.
Because an inhomogeneous Poisson process model is one in which each point location is determined independently of
the other point locations (unlike the other typical models described
above), we adopt an inhomogeneous Poisson process as a spatial point process model for $\{\bfw_i\}_i$. 
In practice, it is impossible for us to reject the internal model based
on the data that we have. However, because of this cross-correlation
analysis and the fact that the high- or low-density subregions are
closely related to the existence of facilities there, we assume the
approximation that the WLAN AP locations follow an inhomogeneous Poisson
process with intensity $\lambda(\bfx)$, where $\lambda(\bfx)$ is the function
of location $\bfx$.  

\OMIT{
Based on the analytical results mentioned above as well as the fact that
some common sparse regions among different operators are regions such as
parks or shrines in which WLAN APs are not allowed, we conclude that the
cause of the clump is the inhomogeneity of the attractiveness of each
location and not the relative locations of other WLAN APs. 
That is, the external models should be chosen, although the internal model can be fit to the given data.
The intensity of the external model should depend on the location of the AP of a WLAN, not on the locations of other WLAN APs.
As a result, we adopt an inhomogeneous Poisson model with intensity $\lambda(\bfx)$ as the spatial point process model for WLAN APs in this paper, where the intensity $\lambda(\bfx)$ is the function of location $\bfx$. 
}

Assuming that $\{\bfw_i\}_i$ follow an inhomogeneous Poisson process, the
later sections analyze, derive, and evaluate performance metrics, such
as mean available bandwidth for a user and the number of vertical
handovers through integral geometry.

\section{Preliminaries for analysis}
\subsection{Notation}
In the remainder of this paper, $\sizex{X}$ denotes the size of  $X$, $\lengthx{X}$ denotes the perimeter length of $X$, $\partial X$ denotes the boundary of $X$, and $\overline{X}$ denotes the complementary set (region) of $X$ for region $X\subset \mathbb{R}^2$.
In addition, for a given line $G$, $\sigma(X)$ denotes the length of the chord $X\cap G$.
For regions $X,Y\subset \mathbb{R}^2$, $\arc{X}{Y}$ denotes the length of the arc created by $\partial X$ and $Y$, that is, the part of $\partial X$ included in $Y$.

\subsection{Integral geometry and geometric probability} 
We introduce the concepts of integral geometry and geometric probability \cite{Santalo} as a preliminary to the following analysis.

Consider a bounded set $X\subset \mathbb{R}^2$ and a condition $X_c$.
A typical example of $X_c$ is $X_c=\{X\cap X_1\neq \emptyset\}$ for a given set $X_1$.
Here, $X$ is, for example, a region covered by a WLAN AP and $X_1$ is the cell of a cellular network.
Integral geometry provides a method for measuring the expectation of the quantity $q(X)$ for the set of positions of $X$ satisfying a condition $X_c$.
Then, if we would like to consider the size of the intersection of $X$ and $X_1$ when $X$ intesects $X_1$, set $q(X)=\sizex{X\cap X_1}$ and $X_c=\{X\cap X_1\neq \emptyset\}$.

For a set of $X$ whose position is defined by the reference point
$(x,y)$ and the angle $\theta$ that a reference line fixed in $X$ makes
with another reference line fixed to the fixed coordinates, integral
geometry defines $E[q(X)]$ by $\int_{X_c} q(X)dX / \int_{X_c} dX=\int_{X_c} q(X)dx\,dy\,d\theta/\int_{X_c} dx\,dy\,d\theta$.
The numerator means the integral of $q(X)$ at a position $(x,y,\theta)$ uniformly over the possible parameter space $(x,y,\theta)$ satisfying $X_c$, and the denominator means the area size of the parameter space $(x,y,\theta)$ satisfying $X_c$.
That is, the numerator is (roughly speaking) a summation of $q(X)$ at every points specified by $(x,y,\theta)$ satisfying $X_c$, and the denominator is (roughly speaking) the number of points satisfying $X_c$.
Therefore, it is an expectation of $q(X)$.

In particular, if $q(X)=\bfone$ (the position $(x,y,\theta)$ satisfies $Y_c$) where $\bfone(\cdot)$ denotes the indicator function and $Y_c\subseteqq X_c$, $E[q(X)]=\int_{Y_c} dx\,dy\,d\theta/\int_{X_c} dx\,dy\,d\theta=\Pr(Y_c|X_c)$ is the (conditional) probability of the positions of $X$ satisfying a condition $Y_c$ among the positions of $X$ satisfying a condition $X_c$.
(This is called a geometric probability \cite{Santalo}.)
In this sense, $\int_{Y_c} dx\,dy\,d\theta$ is a non-normalized probability, because it is proportional to the probability and normalized by $\int_{X_c} dx\,dy\,d\theta$.
In the remainder of this paper, this non-normalized probability is called the measure of the set of positions of $X$ satisfying a condition $Y_c$.

\begin{figure}[htb] 
\begin{center} 
\includegraphics[width=6cm,clip]{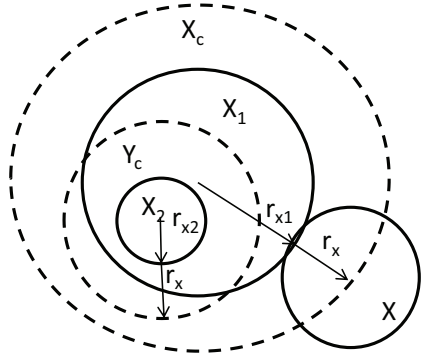} 
\caption{Simple example} 
\label{simpleExample} 
\end{center} 
\end{figure}

A simple example is shown in Fig. \ref{simpleExample}, where $X$ is a disk of radius $r_x$, $X_c=\{X\cap X_1\neq \emptyset\}$, $X_1$ is a disk of radius $r_{x1}$, $Y_c=\{X\cap X_2\neq \emptyset\}$, and $X_2\subset X_1$ is a disk of radius $r_{x2}$.
Because this example is independent of $\theta$, we can easily draw a picture.
Because integral geometry implicitly assumes that the position $(x,y,\theta)$ uniformly moves in the parameter space (if not explicitly indicated otherwise), we can easily understand that $\Pr(Y_c|X_c)$ is given by $\int_{Y_c} dx\,dy\,d\theta/\int_{X_c} dx\,dy\,d\theta=\int_{Y_c} dx\,dy/\int_{X_c} dx\,dy=(r_{x2}+r_x)^2/(r_{x1}+r_x)^2$.

When $X$ is a line $G$, we should use the parameterization by the angle $\theta$, in which the direction perpendicular to $G$ is a fixed direction ($-\pi \leq \theta \leq \pi$), and by its distance $p$ from the origin $O$ ($0\leq p$). 
(We can use another parameterization, but we cannot calculate the integral uniformly over the possible parameter space $X_c$ when the parameters $\theta$ and $p$ are not used.
This is because integral geometry requires the calculated results to be invariant under the group of motions in the plane.)
By using $\theta$ and $p$, the expectation of the quantity $q(G)$ satisfying $X_c$ can be calculated by $\int_{X_c} q(G)dG/\int_{X_c} dG=\int_{X_c} q(G)dp\,d\theta/\int_{X_c} dp\,d\theta$.

In Section \ref{dynamic}, we assume that a user moves on $G$.
By defining $X_c$ such that $G$ intersects the cell and setting $q(G)$ as the chord length of $G$ in the cell, we can calculate the mean distance (that is, the mean length of the chord) that a user moves in the cell.

\subsection{Known basic formulas}
The following basic formulas are known or directly derived through integral geometry.

Accoring to Eqs. (3.12) and (3.6) in \cite{Santalo}, for a fixed convex set $K$, the measure in which the set of positions of a line $G$ that meets $K$ is given by 
\begin{equation}
\int_{\intersect{K}{G}}dG=\lengthx{K},\label{3.12}
\end{equation}
and the (non-normalized) mean length of the chord made by $\partial K$ and $G$ is given by 
\begin{equation}
\int_{\intersect{K}{G}}\sigma(K)dG=\pi\sizex{K}.\label{chordLength}
\end{equation}

For a fixed convex set $K_0$, the measure of the set of positions of a convex set $K_1$ that meets $K_0$ is given as follows (Eq. (6.48) in \cite{Santalo}).
\begin{eqnarray}
\int_{\intersect{K_0}{K_1}}dK_1&=&f(K_0,K_1)\label{6.48}
\end{eqnarray}
where $f(X,Y)\defeq 2\pi(\sizex{X}+\sizex{Y})+\lengthx{X}\cdot\lengthx{Y}$.

Particularly when $K_0$ is a point, Eq. (\ref{6.48}) becomes
\begin{equation}
\int_{\intersect{K_0}{K_1}}dK_1=2\pi \sizex{K_1}.\label{point}
\end{equation}

For a fixed convex set $K_0$, the measure of the set of positions of a convex set $K_1$ that is contained in $K_0$ is given as follows (Eq. (6.52) in \cite{Santalo}).
\begin{eqnarray}
\int_{K_1\subset K_0}dK_1&=&2\pi (\sizex{K_0}+\sizex{K_1})-\lengthx{K_0}\cdot\lengthx{K_1}\label{6.52}
\end{eqnarray}
Formally speaking, additional conditions on the curvature of $\partial K_0$ and that of $\partial K_1$ are needed for Eq. (\ref{6.52}) where $\partial K$ for a set $K\subset \mathbb{R}^2$ means the boundary of $K$.

For a fixed set $K_0$, the integral of $\sizex{K_0\cap K_1}$ over the position of the set of $K_1$ is given as follows. (Eq. (\ref{6.57}) is Eq. (6.57) in \cite{Santalo}. Although Eq. (6.57) in \cite{Santalo} does not include $\pi$, Eqs. (6.55) and (6.56) used in \cite{Santalo} to derive Eq. (6.57) show that Eq. (\ref{6.57}) is correct and the original Eq. (6.57) in \cite{Santalo} is incorrect.).
\begin{eqnarray}
\int_{\intersect{K_0}{K_1}}\sizex{K_0\cap K_1}dK_1&=&2\pi\sizex{K_0}\cdot\sizex{K_1}\label{6.57}
\end{eqnarray}

Due to Eq. (6.61) in \cite{Santalo} and Eq. (\ref{6.52}), Eq. (\ref{6.61}) concerning the integral of $\arc{K_1}{K_0}$ (the length of the arc of $\partial K_1$ intersecting $K_0$) is given as follows.
\begin{eqnarray}
\int_{\intersect{\partial K_0}{K_1}}\arc{K_1}{K_0}dK_1&=&\int_{\intersect{K_0}{K_1}}\arc{K_1}{K_0}dK_1-\int_{K_1\subset K_0}\lengthx{K_1}dK_1\cr
&=&2\pi\sizex{K_0}\cdot\lengthx{K_1}-\lengthx{K_1}(2\pi (\sizex{K_0}+\sizex{K_1})-\lengthx{K_0}\cdot\lengthx{K_1})\cr
&=&\lengthx{K_1}(\lengthx{K_0}\cdot\lengthx{K_1}-2\pi \sizex{K_1})\label{6.61}
\end{eqnarray}
Based on Eq. (\ref{6.57}), for any $K_2\subset K_0$,
\begin{eqnarray}
\int_{\intersect{K_0}{K_1}}\sizex{K_2\cap K_1}dK_1=\int_{\intersect{K_2}{K_1}}\sizex{K_2\cap K_1}dK_1=2 \pi \sizex{K_1}\cdot\sizex{K_2}\label{6.57-2}
\end{eqnarray}
This is because (i) $\int_{\intersect{K_0}{K_1}}\sizex{K_2\cap K_1}dK_1=\int_{\intersect{K_2}{K_1}}\sizex{K_2\cap K_1}dK_1+\int_{K_2\cap K_1=\emptyset,\intersect{K_0}{K_1}}\sizex{K_2\cap K_1}dK_1$, and (ii) the second term is 0 due to $\sizex{K_2\cap K_1}=0$ for $K_2\cap K_1=\emptyset$.

\subsection{Extension of basic formulas}\label{approx_section}
To analyze an inhomogeneous Poisson process, we propose the following extensions for the basic formulas mentioned above.
We assume that $\Omega$ is a convex set and that $dK_1=d\bfw\  d\gamma$ where $\bfw$ is a reference point in $K_1$, and $\gamma$ is the angle characterizing $K_1$.
These extensions require that the reference point $\bfw$ must be within $\Omega$.
For a fixed $K_0$, the relative location of $\Omega$ is assumed to be fixed.
In the following, the term $\hat r(K_1)$ is similar to the radius of $K_1$, and we adopt the definition $\hat r(K_1) \defeq \lengthx{K_1}/(2\pi)$.
This term is identical to the radius when $K_1$ is a disk.

\begin{figure}[htb] 
\begin{center} 
\includegraphics[width=6cm,clip]{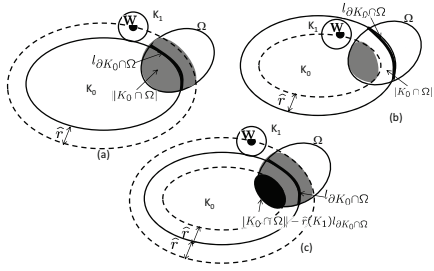} 
\caption{Derivations of approximations} 
\label{approx} 
\end{center} 
\end{figure}

Eqs. (\ref{6.48}) and (\ref{6.52}) are approximately extended to the following.
For a fixed convex set $K_0$, the measure of the set of positions of a convex set $K_1$ that meets $K_0$ and is contained in $K_0$ are given as follows.
\begin{eqnarray}
\int_{\intersect{K_0}{K_1},\bfw\in \Omega}d\bfw\  d\gamma&\approx&2\pi (\eqFirst{K_0}{K_1}{\Omega})\label{6.48r}\\
\int_{K_1\subset K_0,\bfw\in \Omega}d\bfw\  d\gamma&\approx&2\pi (\eqSecond{K_1}{K_0}{\Omega}) \label{6.52r}
\end{eqnarray}
This is due to Fig. \ref{approx}-(a), (b) where the gray regions indicate where $\bfw$ can exist.
Thus, the right-hand sides of Eqs. (\ref{6.48r}) and (\ref{6.52r}) are approximations of the size of these regions times 2$\pi$.

When $K_0$ is a point, the following approximation, which is an analogy of Eq. (\ref{point}), is proposed.  
\begin{eqnarray}
\int_{\intersect{K_0}{K_1},\bfw\in \Omega}d\bfw\  d\gamma&\approx&2\pi \sizex{K_1}\Pr(K_0\in \Omega)\label{6.48rr}
\end{eqnarray}

Eq. (\ref{6.57}) is approximately extended to the following.
For a fixed set $K_0$, the integral of $\sizex{K_0\cap K_1}$ and the integral of $\arc{K_1}{K_0}$ over positions of a set $K_1$ are given as follows.
\begin{eqnarray}
\int_{\intersect{K_0}{K_1},\bfw\in \Omega}\sizex{K_0\cap K_1}d\bfw\  d\gamma&\approx&2\pi \{(\sizex{K_0\cap\Omega}-\hat r(K_1)l_{\partial K_0 \in \Omega})\sizex{K_1}+2\hat r(K_1)l_{\partial K_0 \in \Omega}(\sizex{K_1}/2)\}\cr
&=&2 \pi \sizex{K_1}\cdot\sizex{K_0\cap\Omega}\label{6.57r}
\end{eqnarray}
This is due to Fig. \ref{approx}-(c).
Here, $\sizex{K_0\cap K_1}$ becomes $\sizex{K_1}$ for $\bfw$ in the black region in this figure, and it becomes approximately half of $\sizex{K_1}$ for $\bfw$ in the gray region.
The size of the black region is approximately $\sizex{K_0\cap\Omega}-\hat r(K_1)l_{\partial K_0 \in \Omega}$ and that of the gray region is approximately $2\hat r(K_1)l_{\partial K_0 \in \Omega}$.
Therefore, we obtain Eq. (\ref{6.57r}).

Applying Eq. (\ref{6.57r}), for any $K_2\subset K_0$, 
\OMIT{
\begin{eqnarray}
\int_{\intersect{K_0}{K_1},\bfw\in \Omega}\sizex{K_2\cap K_1}d\bfw\  d\gamma=\int_{\intersect{K_2}{K_1},\bfw\in \Omega}\sizex{K_2\cap K_1}d\bfw\  d\gamma
\approx 2 \pi \sizex{K_1}\cdot\sizex{K_2\cap\Omega}\label{6.57rr}
\end{eqnarray}
}
\begin{eqnarray}
\int\limits_{\intersect{K_0}{K_1},\bfw\in \Omega}\sizex{K_2\cap K_1}d\bfw\  d\gamma=\int_{\intersect{K_2}{K_1},\bfw\in \Omega}\sizex{K_2\cap K_1}d\bfw\  d\gamma
\approx 2 \pi \sizex{K_1}\cdot\sizex{K_2\cap\Omega}\label{6.57rr}
\end{eqnarray}

Eq. (\ref{6.61}) is approximately extended to the following.
Because $\arc{K_1}{K_0}\approx \lengthx{K_1}/2$ and $\int_{\intersect{\partial K_0}{K_1},\bfw\in \Omega}d\bfw\approx 2\hat r(K_1)\arc{K_0}{\Omega}$,
\OMIT{
\begin{eqnarray}
\int_{\intersect{\partial K_0}{K_1},\bfw\in \Omega}\arc{K_1}{K_0}d\bfw\  d\gamma&\approx&2\pi \hat r(K_1)\arc{K_0}{\Omega}\lengthx{K_1}.\label{6.61r}
\end{eqnarray}
}
\begin{eqnarray}
\displaystyle \int\limits_{\intersect{\partial K_0}{K_1},\bfw\in
 \Omega}\arc{K_1}{K_0}d\bfw\  d\gamma \approx 2\pi \hat r(K_1)\arc{K_0}{\Omega}\lengthx{K_1}.\label{6.61r}
\end{eqnarray}

Eqs. (\ref{6.48r}), (\ref{6.52r}), (\ref{6.57r}) (as a result, Eq. (\ref{6.57rr})), and (\ref{6.61r}) are exact when $\partial K_0$ is a line segment in $\Omega$, $K_1$ is a disk with radius $r_{K_1}$, and $\partial \Omega\subset (\overline{K_0}\cap (K_0\oplus r_{K_1}))$ is a line segment vertical to $\partial K_0$ where $K_0\oplus r_{K_1}$ is a Minkoski sum of $K_0$ and a disk with radius $r_{K_1}$.

\section{Model}
We focus on a single cell of a cellular network and WLAN APs around it for the remainder of this paper.
The subregion of this cell can be classified according to the radio channel quality, which identifies the channel quality indicator (CQI) \cite{LTE}, \cite{LTE-A}, \cite{LTEbook}.
Let $C_i$ be a region in the cell where the CQI is $i$.
Let us assume that $C_{n}\subset C_{n-1}\subset \cdots \subset C_{1}=C$ and that $C_{n+1}=\emptyset$ to simplify the notation.
Let $s_i$ be the achieved bitrate (bps) of the radio channel used in $C_i-C_{i+1}$.

Suppose that a public WLAN is provided to offload Internet access traffic.
The $i$-th WLAN AP is located at $\bfw_i$ and has the coverage region $D_i$ with the angle $\gamma_i$ made by a reference line fixed in $D_i$ and a reference line fixed to the fixed coordinates where $\intersect{C}{D_i}$ ($1\leq i\leq l$).
In the remainder of this paper, we use the following assumptions if not explicitly indicated otherwise:
\begin{itemize}
\item $C_i$ and $D_j$ are convex; 
\item The set of locations $\{\bfw_i\}_i$ indepedently follows an inhomogeneous Poisson process with intensity $\lambda(\bfx)$ at $\bfx$; 
\OMIT{
\item $\lambda(\bfx)=\begin{cases}
		      \lambda_H, & \text{for~} \bfx\in \text{high
		      density region~} \Omega_H\subset {\mathbb R}^2,\\
		      \lambda_L, & \text{for~} \bfx\in \text{low density region~} \Omega_L\subset {\mathbb R}^2,\\
		      \lambda_0, & \text{for~} \bfx\not
		      \in\Omega_H,\Omega_L \text{~where~} \lambda_L\leq \lambda_0\leq \lambda_H
		\end{cases}$
}
\item $\lambda(\bfx)=\cases{\lambda_H, & for $\bfx\in$ high density
      region $\Omega_H\subset {\mathbb R}^2$,\cr \lambda_L, & for
      $\bfx\in$ low density region $\Omega_L\subset {\mathbb R}^2$,\cr
      \lambda_0, & for $\bfx\not \in\Omega_H,\Omega_L$}$ where
      $\lambda_L\leq \lambda_0\leq \lambda_H$;  
\item $\Omega_H\cap\Omega_L=\emptyset$; 
\item $\gamma_i$ is uniformly distributed; 
\item When a user can use either a cellular network or WLAN, he/she uses the WLAN;
\item The relative locations of $\Omega_H$ and $\Omega_L$ from $C$ are fixed;
\item Locations of line $G$ are indepepndent of $\{\bfw_i\}_i$.
\end{itemize}
Let $s_w$ be the available bandwidth (bps) of the WLAN, the relative additional intensity $\rho_H\defeq(\lambda_H-\lambda_0)/\lambda_0$,
the relative reduced intensity $\rho_L\defeq(\lambda_0-\lambda_L)/\lambda_0$, the mean intensity $\lambda_C\defeq
\{\lambda_H\sizex{C\cap\Omega_H}+\lambda_L\sizex{C\cap\Omega_L}+\lambda_0(\sizex{C}-\sizex{C\cap\Omega_H}-\sizex{C\cap\Omega_L})\}/\sizex{C}$,
and the relative normal region intensity
$\rho_0\defeq\lambda_0/\lambda_C$.

This paper addresses the mean number $N_h$ of vertical handovers between the cell and a WLAN coverage region and the two types of mean available bandwidth and probability in which a user can use the bandwidth (bps) faster than a certain value as performance metrics.
The first type of mean available bandwidth is called the static available bandwidth $B_s$ (bps), i.e., the bandwidth available by a user who does not move, and the second type is the dynamic available bandwidth, $B_d$ (bps), i.e., the bandwidth available by a user who moves.
$B_s$ and $B_d$ are the main performance metrics from the user point of view.
On the other hand, $N_h$ is important mainly from a network operator point of view because additional tasks are required in the network at the handover.
For metric $x$, $\tilde x$ denotes $x$ without introducing a public WLAN.

The mean total throughput $T_d$ is defined by the sum of bits that a user moving on a line $G$ ($\intersect{C}{G}$) at a unit speed in the cell can send by making full use of the radio channel if no other competing users, and it is averaged over various positions of $G$.
$B_s$ is the mean available bandwidth at which users distributed uniformly over $C$ can use the radio channel if no other competing users, and $B_d$ is the ratio of the mean total throughput $T_d$ to the mean total sojourn time within the cell.
The total sojourn time is defined by the sum of times during which a user moving on a line $G$ ($\intersect{C}{G}$) at a unit speed stays in the cell and it is averaged over various positions of $G$.
In addition, let $q_s(x)$ ($q_d(x)$) be the probability that a user staying somewhere in $C$ (moving on a line $G$ at a unit speed) can use the bandwidth (bps) faster than $x$.

Throughout the analysis, the times for signaling and its processing, the influence of other competing users, and the interference between WLAN APs are not considered.
Therefore, this analysis provides ideal performance.

\section{Theoretical analysis without introducing WLANs}
This section derives the performance metrics when WLANs are not introduced.

Because the probability that a point randomly chosen in $C$ is in $C_{i}-C_{i+1}$ is $(\sizex{C_{i}}-\sizex{C_{i+1}})/\sizex{C}$, we obtain the following result.
\begin{result}
The mean and cumulative probabilistic distribution of the bandwidth available by the user who does not move are given as follows.
\begin{eqnarray}
\tilde B_s&=&\frac{\sum_{i=1}^n(\sizex{C_{i}}-\sizex{C_{i+1}})s_{i}}{\sizex{C}}\label{BsNoLAN}\\
\tilde q_s(x)&=& \frac{\sum_{i=1}^n(\sizex{C_{i}}-\sizex{C_{i+1}})\bfone(s_i\geq x)}{\sizex{C}}\label{qsNoLAN}
\end{eqnarray}
\end{result}

Consider a line $G(\theta,p)$.
Assume that a tagged user moves on $G(p,\theta)$ at a unit speed.
\begin{result}
\begin{eqnarray}
\tilde B_d&\defeq& \frac{\tilde T_d}{\int_{\intersect{C}{G}}\sigma(C)dG/\int_{\intersect{C}{G}}dG}=\tilde B_s,\label{BdNoLAN}\\
\tilde q_d(x)&\defeq&\frac{\int_{\intersect{C}{G}}\sum_{i=1}^n \sigma(C_i-C_{i+1})\bfone(s_i\geq x)dG}{\int_{\intersect{C}{G}}\sigma(C)dG}=\tilde q_s(x),\label{qdNoLAN}\\
\tilde N_h&=&0.
\end{eqnarray}
Here, the mean total throughput $\tilde T_d$ is given as follows: $\tilde T_d\defeq\frac{\int_{\intersect{C}{G}}\sum_{i=1}^n s_i\sigma(C_i-C_{i+1})dG}{\int_{\intersect{C}{G}}dG}=\sum_{i=1}^n\pi(\sizex{C_{i}}-\sizex{C_{i+1}})s_{i}/\lengthx{C}.$
\end{result}

\proof
The length $\sigma(C_i-C_{i+1})$ of the chord $G(p,\theta)\cap(C_i-C_{i+1})$ is given by the following equation due to Eq. (\ref{chordLength}), for $1\leq i\leq n$.
\begin{eqnarray}
\int_{\intersect{C}{G}}\sigma(C_i-C_{i+1})dG
&=&\int_{\intersect{C_i}{G}}\sigma(C_i)dG-\int_{\intersect{C_{i+1}}{G}}\sigma(C_{i+1})dG=\pi  (\sizex{C_i}-\sizex{C_{i+1}} )
\end{eqnarray}
Because we consider the set of lines $\{\intersect{C}{G}\}$, we need to normalize by $\int_{\intersect{C}{G}}dG$.
Therefore, using Eq. (\ref{3.12}), we obtain $\tilde T_d$.

Because the mean length of the chord $G\cap C$ is $\int_{\intersect{C}{G}}\sigma(C)dG/\int_{\intersect{C}{G}}dG=\pi  \sizex{C}/\lengthx{C}$, we obtain Eqs. (\ref{BdNoLAN}) and (\ref{qdNoLAN}).

Because there are no WLANs, $\tilde N_h=0$.
$\square$

\section{Theoretical analysis after introducing WLANs}
This section derives the performance metrics when WLANs are introduced.
Because it seems impossible to derive explicit formulas for the performance metrics under an inhomogeneous Poisson process, they are approximation formulas.
However, they become exact under a homogeneous Poisson process.

In the remaining, we use the notation $C_x\cap_{i=i_1,\cdots,i_m}D_i\defeq C_x\cap D_{i_1}\cap D_{i_2}\cap\cdots\cap D_{i_m}$, $\{\intersect{C_x}{D_i}\}_{i=1}^l\defeq\{\intersect{C_x}{D_1},\cdots,\intersect{C_x}{D_l} \}$, and $(dD)^l\defeq dD_1\ \cdots\ dD_l$.
Here, $C_x$ is $C$ or $C_i$, ($i=1,\cdots,l$).

At the beginning, we provide the following result, which is often used in the remaining in this section.
\begin{result}
Under a homogeneous Poisson process,
\begin{eqnarray}
&&\int_{\{\intersect{C}{D_i}\}_{i=1}^l}(dD)^l=\Pi_{i=1}^l f(C,D_i).\label{homo-1}
\end{eqnarray}
Under an inhomogeneous Poisson process,
\begin{eqnarray}
&&\int_{\{\intersect{C}{D_i}\}_{i=1}^l}(dD)^l\approx\rho_0^l\Pi_{i=1}^l g_1(C,D_i)\label{pj1}
\end{eqnarray}
where $g_1(C,D_i)\defeq f(C,D_i)+2\pi (\rho_H (\eqFirst{C}{D_i}{\Omega_H})- \rho_L(\eqFirst{C}{D_i}{\Omega_L}))$.
\end{result}

\proof
Under a homogeneous Poisson process, by repeatedly applying Eq. (\ref{6.48}), we obain Eq. (\ref{homo-1}).

For an inhomogeneous Poisson process, $\lambda_H$ ($\lambda_L$, $\lambda_0$) is the intensity that a WLAN internet AP is located in the high (low, normal) density region.  
Therefore, for an arbitrary functon $\phi$ and any condition $X_c$, 
\begin{eqnarray*}
\lambda_C\int_{X_c}\phi\,  dD_i&= &\lambda_0\int_{X_c,\bfw_i\in \overline{\Omega_H\cup\Omega_L}}\phi\,  d\bfw_i\  d\gamma_i +\lambda_H\int_{X_c,\bfw_i \in \Omega_H}\phi\,  d\bfw_i\  d\gamma_i+\lambda_L\int_{X_c,\bfw_i \in \Omega_L}\phi\,  d\bfw_i\  d\gamma_i.
\end{eqnarray*}
As a result,
\begin{eqnarray}
\int_{X_c}\phi\,  dD_i&=&\rho_0\{\int_{X_c,\bfw_i\in {\mathbb R}^2}\phi\,  d\bfw_i\  d\gamma_i +\rho_H\int_{X_c,\bfw_i \in \Omega_H}\phi\,  d\bfw_i\  d\gamma_i- \rho_L\int_{X_c,\bfw_i \in \Omega_L}\phi\,  d\bfw_i\  d\gamma_i\}.\label{inhomo0}
\end{eqnarray}
Set $X_c=\intersect{C}{D_i}$ and $\phi=1$, and apply Eq. (\ref{inhomo0}).  
Because of Eqs. (\ref{6.48}) and (\ref{6.48r}), 
\begin{eqnarray}
\int_{\intersect{C}{D_i}}dD_i
&=&\rho_0\{\int_{\intersect{C}{D_i},\bfw_i\in {\mathbb R}^2}d\bfw_i\  d\gamma_i +\rho_H\int_{\intersect{C}{D_i},\bfw_i \in \Omega_H}d\bfw_i\  d\gamma_i- \rho_L\int_{\intersect{C}{D_i},\bfw_i \in \Omega_L}d\bfw_i\  d\gamma_i\}\cr
&\approx&\rho_0g_1(C,D_i).\label{dDi}
\end{eqnarray}
$\square$

\subsection{Derivation of $B_s$ and $q_s(x)$}
This subsection provides $B_s$ and $q_s(x)$.
First, we evaluate the expected area size of $C_j\cap_{m=i_1,\cdots,i_k}D_m$ under the condition $\{\intersect{C}{D_i}\}_{i=1}^l$.
Second, by describing $p_j$, which is the probability that the point is in $C_j-C_{j+1}$ and is covered by at least a single WLAN, with this expected area size, we derive $p_j$.
Third, based on $p_j$, $B_s$ and $q_s(x)$ are derived.

\subsubsection{Derivation of $E[\sizex{C_j\cap_{m=i_1,\cdots,i_k}D_m}\ |\{\intersect{C}{D_i}\}_{i=1}^l]$}
Let $E[\sizex{C_j\cap_{m=i_1,\cdots,i_k}D_m}\ |\{\intersect{C}{D_i}\}_{i=1}^l]$ be the expected area size of $C_j\cap_{m=i_1,\cdots,i_k}D_m $ under the condition $\{\intersect{C}{D_i}\}_{i=1}^l$.

This expected area size is given by the following.
\begin{result}\label{size-result}
\begin{eqnarray}
E[\sizex{C_j\cap_{m=i_1,\cdots,i_k}D_m}\ |\{\intersect{C}{D_i}\}_{i=1}^l]
&\defeq&\frac{\int_{\{\intersect{C}{D_i}\}_{i=1}^l}\sizex{C_j\cap_{m=i_1,\cdots,i_k}D_m }(dD)^l}{\int_{\{\intersect{C}{D_i}\}_{i=1}^l}(dD)^l}\label{size-def}
\end{eqnarray}
The denominator is given by Eq. (\ref{homo-1}) or (\ref{pj1}).
The numerator is given by Eq. (\ref{homo-0}) or (\ref{g2}) shown below.

For a homogeneous Poisson process
\begin{eqnarray}
&&\int_{\{\intersect{C}{D_i}\}_{i=1}^l}\sizex{C_j\cap_{m=i_1,\cdots,i_k}D_m }(dD)^l\cr
&=&\sizex{C_j}\Pi_{i\neq i_1,\cdots,i_k} f(C,D_i)\cdot\Pi_{j= i_1,\cdots,i_k}(2\pi\sizex{D_{j}}).\label{homo-0}
\end{eqnarray}
For an inhomogeneous Poisson process,
\begin{eqnarray}
&&\int_{\{\intersect{C}{D_i}\}_{i=1}^l}\sizex{C_j\cap_{m=i_1,\cdots,i_{k}}D_m }(dD)^l\cr
&\approx&\rho_0^lg_2(C_j,k)\Pi_{i\neq i_1,\cdots,i_k} g_1(C,D_i)\cdot\Pi_{m= i_1,\cdots,i_k}(2\pi\sizex{D_{m}})\label{g2}
\end{eqnarray}
where $g_2(C_j,k)\defeq \sizex{C_j}+\sizex{C_j\cap\Omega_H}((1+\rho_H)^k-1) +\sizex{C_j\cap\Omega_L}((1-\rho_L)^k-1)$.
\end{result}

\proof
For a homogeneous Poisson process, repeatedly apply Eqs. (\ref{6.48}) and (\ref{6.57-2}) to obtain Eq. (\ref{homo-0}).
\begin{eqnarray*}
&&\int_{\{\intersect{C}{D_i}\}_{i=1}^l}\sizex{C_j\cap_{m=i_1,\cdots,i_k}D_m }(dD)^l\cr
&=&\Pi_{i\neq i_1,\cdots,i_k} f(C,D_i)\int_{\intersect{C}{D_{i_1}},\cdots,\intersect{C}{D_{i_k}}}\sizex{C_j\cap_{m=i_1,\cdots,i_k}D_m }dD_{i_1}\cdots dD_{i_k}\cr
&=&(2\pi\sizex{D_{i_k}})\Pi_{i\neq i_1,\cdots,i_k} f(C,D_i)\int_{\intersect{C}{D_{i_1}},\cdots,\intersect{C}{D_{i_{k-1}}}}\sizex{C_j\cap_{m=i_1,\cdots,i_{k-1}}D_m }dD_{i_1}\cdots dD_{i_{k-1}}\cr
&=&\sizex{C_j}\Pi_{i\neq i_1,\cdots,i_k} f(C,D_i)\cdot\Pi_{j= i_1,\cdots,i_k}(2\pi\sizex{D_{j}}).
\end{eqnarray*}

The proof for an inhomogeneous Poisson process is in Appendix \ref{ap2}.$\square$

\subsubsection{Derivation of $p_j$}
In this subsection, we derive the probability that a randomly chosen point is in $C_j-C_{j+1}$ and is covered by at least a single WLAN.
By using this probability, the performance metrics are derived later.

For a randomly chosen point $\bfx\in C$, let $p_j$ be the probability that the point is in $C_j-C_{j+1}$ and is covered by at least a single WLAN.
According to the definition of a geometric probability,
\begin{eqnarray}
p_j&\defeq& \Pr(\bfx\in C_j -C_{j+1},\bfx\in \bigcup_{i=1}^l D_i, \intersect{D_i}{C}, i=1,\cdots,l\vert \bfx\in C,\intersect{D_i}{C}, i=1,\cdots,l)\cr
&=&\frac{\int_{\bfx\in C_j -C_{j+1},\bfx\in \bigcup_{i=1}^l D_i, \{\intersect{C}{D_i}\}_{i=1}^l}d\bfx \  (dD)^l}{\int_{\bfx\in C,\{\intersect{C}{D_i}\}_{i=1}^l}d\bfx \  (dD)^l}.
\end{eqnarray}

By using $E[\sizex{C_j\cap_{m=i_1,\cdots,i_k}D_m}\ |\{\intersect{C}{D_i}\}_{i=1}^l]$, we can describe $p_j$.
\begin{result}
\begin{eqnarray}
p_j&=&\frac{1}{\sizex{C}}\sum_{m=1}^l\sum_{1\leq i_1<\cdots<i_m\leq l}(-1)^{m-1}(E[\sizex{C_j\cap_{i=i_1,\cdots,i_m}D_i}\ |\{\intersect{C}{D_i}\}_{i=1}^l]\cr
&&\qquad \qquad \qquad \qquad \qquad \qquad -E[\sizex{C_{j+1}\cap_{i=i_1,\cdots,i_m}D_i}\ |\{\intersect{C}{D_i}\}_{i=1}^l]).\label{size-pj}
\end{eqnarray}
\end{result}

\proof
Due to Eq. (\ref{point}),
\begin{eqnarray}
p_j&=&\frac{\int_{\{\intersect{C}{D_i}\}_{i=1}^l}\sizex{(C_j -C_{j+1})\cap ( \bigcup_{i=1}^l D_i)}(dD)^l}{\sizex{C}\int_{\{\intersect{C}{D_i}\}_{i=1}^l}(dD)^l}.\label{pj}
\end{eqnarray}
Apply the following equations to the numerator.
\begin{eqnarray}
\sizex{(C_j -C_{j+1})\cap ( \bigcup_{i=1}^l D_i)}&=&\sizex{\bigcup_{i=1}^l (C_j \cap D_i)}-\sizex{\bigcup_{i=1}^l (C_{j+1}\cap D_i)}\label{c_j}\\
\sizex{\bigcup_{i=1}^l (C_j \cap D_i)}&=&\sum_{i=1}^l \sizex{C_j \cap D_i}-\sum_{1\leq i_1<i_2\leq l}\sizex{C_j \cap D_{i_1}\cap D_{i_2}}+\cdots,\label{c_jd}
\end{eqnarray}
\begin{eqnarray}
&&\int_{\{\intersect{C}{D_i}\}_{i=1}^l}\sizex{(C_j -C_{j+1})\cap ( \bigcup_{i=1}^l D_i)}(dD)^l\cr
&=&\int_{\{\intersect{C}{D_i}\}_{i=1}^l}\sum_{m=1}^l\sum_{1\leq i_1<\cdots<i_m\leq l}(-1)^{m-1} \sizex{C_j\cap_{i=i_1,\cdots,i_m}D_i}
-\sizex{C_{j+1}\cap_{i=i_1,\cdots,i_m}D_i}(dD)^l .\label{pj-2}
\end{eqnarray}
Apply the definition by Eq.(\ref{size-def}) to the equation above, we obtain Eq. (\ref{size-pj}). $\square$

Eq. (\ref{size-pj}) is intuitive. 
Because the point $\bfx$ is randmly chosen, $p_j$ is proportional to the area size of $(C_j -C_{j+1})\cap ( \bigcup_{i=1}^l D_i)$.
By taking into account the overlaps, we obtain Eq. (\ref{pj-2}), which is essentially the same with Eq. (\ref{size-pj}).

We are now in the postion to describe $p_j$.
\begin{result}
Under an inhomogeneous Poisson process, $p_j$ is approximately given by 
\begin{eqnarray}
p_j&\approx&\frac{\sum_{m=1}^l (-1)^{m-1}(g_2(C_j,m)-g_2(C_{j+1},m))b_h(m|C)}{\sizex{C}}\label{pj_result}\cr
&&
\end{eqnarray}
where 
$b_h(m|C)\defeq\sum_{1\leq i_1<\cdots<i_m\leq l} $ $\Pi_{i=i_1,\cdots,i_m}(2\pi\sizex{D_{i}}/g_1(C,D_i))$.

For $j=1,\cdots, n$, when $\sizex{D_i}=\sizex{D}$ and $\lengthx{D_i}=\lengthx{D}$ for all $i$, Eq. (\ref{pj_result}) becomes 
\begin{eqnarray}
p_j&\approx&\frac{(1-b_0)g_0(j)+(1-b_H)g_H(j)+(1-b_L)g_L(j)}{\sizex{C}} \label{sameD}
\end{eqnarray}
where $g_H(j)\defeq \sizex{C_j\cap\Omega_H}-\sizex{C_{j+1}\cap\Omega_H}$, $g_L(j)\defeq \sizex{C_j\cap\Omega_L}-\sizex{C_{j+1}\cap\Omega_L}$, 
$g_0(j)\defeq\sizex{C_j-C_{j+1}}-g_H(j)-g_L(j)$, $b_0\defeq(1-\frac{2\pi\sizex{D}}{g_1(C,D)})^l$, $b_H\defeq(1-\frac{2\pi(1+\rho_H)\sizex{D}}{g_1(C,D)})^l$ and $b_L\defeq(1-\frac{2\pi(1-\rho_L)\sizex{D}}{g_1(C,D)})^l$.

Under a homegneous Poisson process, this approximation formula becomes exact and simplified into 
\begin{eqnarray}
p_j&=&\frac{\sizex{C_j-C_{j+1}}}{\sizex{C}}\{1-\Pi_{i=1}^l(1-\frac{2\pi\sizex{D_i}}{f(C,D_i)})\}.\label{homo}
\end{eqnarray}
\end{result}

\noindent {\bf Proof of Eq. (\ref{pj_result}):}

Apply Eqs. (\ref{pj1}) and (\ref{g2}) to Eq. (\ref{size-def}).
\begin{eqnarray}
&&E[\sizex{C_j\cap_{m=i_1,\cdots,i_k}D_m}\ |\{\intersect{C}{D_i}\}_{i=1}^l]\cr
&\approx&g_2(C_j,k)\Pi_{m= i_1,\cdots,i_k}(2\pi\sizex{D_{m}}/g_1(C,D_m))
\end{eqnarray}
According to Eq. (\ref{size-pj}), 
\begin{eqnarray}
p_j&\approx&\frac{1}{\sizex{C}}\sum_{m=1}^l\sum_{1\leq i_1<\cdots<i_m\leq l}(-1)^{m-1}(g_2(C_j,m)-g_2(C_{j+1},m))\Pi_{k= i_1,\cdots,i_m}(2\pi\sizex{D_k}/g_1(C,D_k))\cr
&=&\frac{1}{\sizex{C}}\sum_{m=1}^l(-1)^{m-1}(g_2(C_j,m)-g_2(C_{j+1},m))b_h(m|C).
\end{eqnarray}
$\square$

\noindent {\bf Proof of Eq. (\ref{sameD}):}
Note $b_h(m|C)={}_lC_m(2\pi\sizex{D}/g_1(C,D))^m$ and $g_2(C_j,m)-g_2(C_{j+1},m)=g_0(j)+(1+\rho_H)^m g_H(j)+(1-\rho_L)^m g_L(j)$. 
Because $\sum_{m=1}^l(-1)^{m-1}{}_lC_m(2\pi\sizex{D}/g_1(C,D))^m=1-(1-2\pi\sizex{D}/g_1(C,D))^l$, $\sum_{m=1}^l(-1)^{m-1}{}_lC_m(2\pi\sizex{D}(1+\rho_H)/g_1(C,D))^m=1-(1-2\pi\sizex{D}(1+\rho_H)/g_1(C,D))^l$, and $\sum_{m=1}^l(-1)^{m-1}{}_lC_m(2\pi\sizex{D}(1-\rho_L)/g_1(C,D))^m=1-(1-2\pi\sizex{D}(1-\rho_L)/g_1(C,D))^l$, Eq. (\ref{sameD}) is derived. 
$\square$

\noindent {\bf Proof of Eq. (\ref{homo}):}
For a homogeneous Poisson process, apply Eqs. (\ref{size-def}), (\ref{homo-1}), and (\ref{homo-0}) to Eq. (\ref{size-pj}).
\begin{eqnarray}
p_j&=&\frac{1}{\sizex{C}}\sum_{m=1}^l\sum_{1\leq i_1<\cdots<i_m\leq l}(-1)^{m-1}(\sizex{C_j}-\sizex{C_{j+1}})\Pi_{j= i_1,\cdots,i_m}(2\pi\sizex{D_{j}}/f(C,D_i))\cr
&=&\frac{\sizex{C_j}-\sizex{C_{j+1}}}{\sizex{C}}\{1-\Pi_{i=1}^l(1-\frac{2\pi\sizex{D_i}}{f(C,D_i)})\}.
\end{eqnarray}
$\square$

Under a homogeneous Poisson process, $\Omega_H=\Omega_L=\emptyset$ and $\rho_H=\rho_L=0$.
Thus,  $g_1(C,D_i)= f(C,D_i)$ and $g_2(C_j,k)=\sizex{C_j}$.
Then, $p_j$ given by Eq. (\ref{pj_result}) becomes identical to that given by Eq. (\ref{homo}).
That is, the approximation formula Eq. (\ref{pj_result}) is exact under a homogeneous Poisson process.

The meaning of Eq. (\ref{homo}) is as follows.
Because the probability that a point $\bfx$ in $C$ is covered by $D_i$ when $\intersect{C}{D_i}$ is $\frac{\int_{\intersect{\bfx}{D_i}}dD_i}{\int_{\intersect{C}{D_i}}dD_i}=\frac{2\pi\sizex{D_i}}{f(C,D_i)}$ due to Eqs. (\ref{6.48}) and (\ref{point}), $\{1-\Pi_{i=1}^l(1-\frac{2\pi\sizex{D_i}}{f(C,D_i)})\}$ indicates the probability that a point in $C$ will be covered by at least one of $D_1,\cdots,D_l$ when they are independently deployed.
Of course, $\frac{\sizex{C_j-C_{j+1}}}{\sizex{C}}$ is the probability that a point in $C$ is in $C_j-C_{j+1}$.

Similarly, we can consider the meaning of Eq. (\ref{pj_result}).
When $\rho_0=1$, $b_h(m|C)$ is approximately the probability that a point in $C$ is covered by $m$ of $D_1,\cdots,D_l$ because $\frac{2\pi\sizex{D_{i}}}{g_1(C,D_i)}$ is approximately the probability that a point in $C$ is covered by $D_i$ intersecting $C$.
In addition, $g_2(C_j,m)$ is the size of $C_j$ modified by the clustered effect because of $C_j$ intersecting $m$ of $D_1,\cdots,D_l$.
Thus, $(g_2(C_j,m)-g_2(C_{j+1},m))b_h(m|C)/\sizex{C}$ is approximately the probability that a point in $C$ is in $C_j-C_{j+1}$ and covered by $m$ of $D_1,\cdots,D_l$ intersecting $C$.
The event in which a point is covered by at least a single WLAN is identical to the occurrence of events in which a point is covered by an individual WLAN.
However, these events are not exclusive.
Therefore, the term $\sum_{m=1}^l (-1)^{m-1}$ taking account of the overlaps of these events appears.

\subsubsection{Derivation of $B_s$ and $q_s(x)$ based on $p_j$}
By using the result mentioned above, we can derive $B_s$ and $q_s(x)$.

\begin{result}
Under an inhomogeneous Poisson process, $B_s$ and $q_s(x)$ (the mean and the cumulative probabilisitic distribution of the available bandwidth that the user who does not move) are approximately given by
\begin{eqnarray}
B_s&\approx&\sum_{j=1}^n [\frac{\sizex{C_{j}}-\sizex{C_{j+1}}}{\sizex{C}}s_j+\frac{s_w-s_j}{\sizex{C}}\sum_{m=1}^l (-1)^{m-1}(g_2(C_j,m)-g_2(C_{j+1},m))b_h(m|C)],\label{inhomoBs}\cr
&&\\
q_s(x)&\approx&\sum_{j=1}^n [\frac{\sizex{C_{j}}-\sizex{C_{j+1}}}{\sizex{C}}\bfone(s_j\geq x)\cr
&&+\frac{\bfone(s_w\geq x)-\bfone(s_j\geq x)}{\sizex{C}}\sum_{m=1}^l (-1)^{m-1}(g_2(C_j,m)-g_2(C_{j+1},m))b_h(m|C)].\label{inhomoqs}
\end{eqnarray}
Under a homogeneous Poisson process, these approximation formulas become exact and simplified into 
\begin{eqnarray}
B_s&=&s_w-\sum_{j=1}^n \frac{(s_w-s_j)\sizex{C_j-C_{j+1}}}{\sizex{C}}\Pi_{i=1}^l(1-\frac{2\pi\sizex{D_i}}{f(C,D_i)})\label{homoBs}\\
q_s(x)&=&\sum_{j=1}^n \frac{\sizex{C_{j}}-\sizex{C_{j+1}}}{\sizex{C}}[\bfone(s_w\geq x)\{1-\Pi_{i=1}^l(1-\frac{2\pi\sizex{D_i}}{f(C,D_i)})\}+\bfone(s_j\geq x)\Pi_{i=1}^l(1-\frac{2\pi\sizex{D_i}}{f(C,D_i)})].\label{homoqs}
\end{eqnarray}
\end{result}

\proof
Because the probability that the point $\bfx$ is in $C_j-C_{j+1}$ but is not covered by any WLANs is $\frac{\sizex{C_i-C_{i+1}}}{\sizex{C}}-p_j$,
\begin{eqnarray}
B_s&=&\sum_{j=1}^n (p_j s_w+(\frac{\sizex{C_{j}}-\sizex{C_{j+1}}}{\sizex{C}}-p_j)s_j),\label{bs1}\\
q_s(x)&=&\sum_{j=1}^n (p_j \bfone(s_w\geq x) +(\frac{\sizex{C_{j}}-\sizex{C_{j+1}}}{\sizex{C}}-p_j) \bfone(s_j\geq x)).\label{qs1}
\end{eqnarray}

Under a homogeneous Poisson process, because $p_j$ is given by Eq. (\ref{homo}), $B_s$ and $q_s(x)$ are given by Eqs. (\ref{homoBs}) and (\ref{homoqs}).

Under an inhomogeneous Poisson process, due to Eq. (\ref{pj_result}), $B_s$ and $q_s(x)$ are given by Eqs. (\ref{inhomoBs}) and (\ref{inhomoqs}).

$\square$

In Eq. (\ref{inhomoBs}), the first term $\frac{\sizex{C_{j}}-\sizex{C_{j+1}}}{\sizex{C}}s_j$ means that the bit rate $s_j$ is available with the probability that a point in $C$ is located in $C_j-C_{j+1}$, and the second term means that the additional bit rate $s_w-s_j$ is available with probability $p_j$.
Eq. (\ref{homoBs}) means that a WLAN bit rate $s_w$ is available but the bit rate may be reduced to $s_j$ with the probability that a point in $C$ is included in $C_j-C_{j+1}$ and not covered by any WLANs.

Similarly to Eq. (\ref{sameD}), Eqs. (\ref{inhomoBs}) and (\ref{inhomoqs}) can be simplfied when $\sizex{D_i}=\sizex{D}$ and $\lengthx{D_i}=\lengthx{D}$ for all $i$.

\subsection{Derivation of $B_d$ and $q_d(x)$}\label{dynamic}
For each $G$, define $\sigma_w\defeq  \sigma(C\cap \bigcup_{i=1}^l D_i)$ and $\sigma_i\defeq\sigma((C_i-C_{i+1})\cap\overline{\bigcup_{j=1}^l D_j})$.
Here, $\sigma_w$ ($\sigma_i$) is a part of $G$ where a user moving on $G$ can use the WLAN (the cellular network with achieved bitrate $s_i$).

This subsection evaluates the mean dynamic available bandwidth $B_d$ and its cumulative probabilistic distribution $q_d(x)$ defined below.
\begin{eqnarray}
B_d \defeq T_d \frac{\int_{\intersect{C}{G}}dG}{\int_{\intersect{C}{G}}\sigma(C)dG}=\frac{T_d\lengthx{C}}{\pi\sizex{C}}\label{def_bd}\\
q_d(x) \defeq \frac{\int\limits_{\intersect{C}{G}, \{\intersect{C}{D_i}\}_{i=1}^l}(\bfone(s_w\geq x) \sigma_w+\sum_{i=1}^n \bfone(s_i\geq x)\sigma_i )dG\  (dD)^l}{\int_{\intersect{C}{G}, \{\intersect{C}{D_i}\}_{i=1}^l}\sigma(C)dG\  (dD)^l}\label{def_qd}
\end{eqnarray}
Here, the mean total throughput $T_d$ is defined by $T_d\defeq\frac{\int_{\intersect{C}{G}, \{\intersect{C}{D_i}\}_{i=1}^l }(s_w\sigma_w+\sum_{i=1}^n s_i\sigma_i )dG\  (dD)^l}{\int_{\intersect{C}{G}, \{\intersect{C}{D_i}\}_{i=1}^l }dG\  (dD)^l}.$

First, the expected length of $\sigma_w$ and that of $\sigma_i$ are described by $E[\sizex{C_j\cap_{m=i_1,\cdots,i_k}D_m}\ |\{\intersect{C}{D_i}\}_{i=1}^l]$.
Second, the mean total throughput $T_d$ is derived based on them.
Third, $B_d$ and $q_d(x)$ are derived based on $T_d$.

\subsubsection{Describing the expected chord length with the condition $\intersect{C}{G}, \{\intersect{C}{D_i}\}_{i=1}^l$}
The expected $\sigma_w$ and $\sigma_i$ with the condition $\intersect{C}{G}, \{\intersect{C}{D_i}\}_{i=1}^l$ are defined as follows.
\begin{eqnarray}
&&E[\sigma_x|\intersect{C}{G}, \{\intersect{C}{D_i}\}_{i=1}^l]\cr
&\defeq&\frac{\int_{\intersect{C}{G}, \{\intersect{C}{D_i}\}_{i=1}^l}\sigma_x dG\  (dD)^l}{\int_{\intersect{C}{G}, \{\intersect{C}{D_i}\}_{i=1}^l}dG\  (dD)^l}\label{chord0}
\end{eqnarray}
where $\sigma_x=\sigma_w$ or $\sigma_i$.
To my surprise, the conditional expectations of $\sigma_w$ and $\sigma_i$ can be described by the conditional expectation of $\sizex{C_j\cap_{m=i_1,\cdots,i_k}D_m}$.

\begin{result}
\begin{eqnarray}
&&E[\sigma_w|\intersect{C}{G}, \{\intersect{C}{D_i}\}_{i=1}^l]\cr
&=&\pi \sum_{m=1}^l (-1)^{m-1}\sum_{1\leq i_1<\cdots<i_m\leq l} E[\sizex{C\cap_{k=i_1,\cdots,i_m}D_k}\ |\{\intersect{C}{D_i}\}_{i=1}^l]/\lengthx{C},\label{sigma_w0}\\
&&E[\sigma_i|\intersect{C}{G}, \{\intersect{C}{D_i}\}_{i=1}^l]\cr
&=&\pi\sum_{m=0}^l (-1)^{m}\sum_{1\leq i_1<\cdots<i_m\leq l} E[\sizex{C_i\cap_{k=i_1,\cdots,i_m}D_k}-\sizex{C_{i+1}\cap_{k=i_1,\cdots,i_m}D_k}\ |\{\intersect{C}{D_j}\}_{j=1}^l]/\lengthx{C}.\label{sigma_i0}
\end{eqnarray}
\end{result}

\noindent {\bf Proof of Eq. (\ref{sigma_w0}):}
According to the definition of $\sigma_w$,
\begin{eqnarray}
\int_{\intersect{C}{G}}\sigma_w dG
&=&\int_{\intersect{C}{G}}\{\sum_{i=1}^l \sigma(C\cap D_i) - \sum_{1\leq i_1<i_2\leq l} \sigma (C\cap D_{i_1}\cap D_{i_2}) + \cdots \}dG.
\end{eqnarray}
Note that $C\cap D_{i_1}\cap D_{i_2} \cap \cdots$ is convex.
Hence, by using Eq. (\ref{chordLength}), we obtain
\begin{eqnarray}
&&\int_{\intersect{C}{G}}\sigma(C\cap_{i=i_1,i_2,\cdots}D_i )dG\cr
&=&\int_{\intersect{C\cap D_{i_1}\cap D_{i_2} \cap \cdots}{G}}  \sigma(C\cap_{i=i_1,i_2,\cdots}D_i) dp\, d\theta\cr
&=&\pi \sizex{C\cap_{i=i_1,i_2,\cdots}D_i}.\label{sigma_0}
\end{eqnarray}
Thus, 
\begin{eqnarray}
&&\int_{\intersect{C}{G}, \{\intersect{C}{D_i}\}_{i=1}^l}\sigma_w dG\  (dD)^l\cr
&=&\pi \int_{\{\intersect{C}{D_i}\}_{i=1}^l}\{\sum_{i=1}^l \sizex{C\cap D_{i}} - \sum_{1\leq i_1<i_2\leq l} \sizex{C\cap D_{i_1}\cap D_{i_2}} + \cdots\}\  (dD)^l.
\end{eqnarray}
On the other hand, the denominator of Eq. (\ref{chord0}) is $\lengthx{C}\int_{\{\intersect{C}{D_i}\}_{i=1}^l}(dD)^l$ because of Eq. (\ref{3.12}).
Due to Eq. (\ref{size-def}), we obtain Eq.(\ref{sigma_w0}). $\square$

\noindent {\bf Proof of Eq. (\ref{sigma_i0}):}
\begin{eqnarray}
\int_{\intersect{C}{G}}\sigma_i dG
&=&\int_{\intersect{C}{G}}\{\sigma(C_i\cap\overline{\bigcup_{j=1}^l D_j})-\sigma(C_{i+1}\cap\overline{\bigcup_{j=1}^l D_j})\}dG\cr
&=&\int_{\intersect{C}{G}}\{\sigma(C_i)-\sigma(C_i\cap\bigcup_{j=1}^l D_j)-\sigma(C_{i+1})+\sigma(C_{i+1}\cap\bigcup_{j=1}^l D_j)\}dG\cr
&=&\int_{\intersect{C}{G}}\{\sigma(C_i)-\sigma(\bigcup_{j=1}^l (C_i\cap D_j))-\sigma(C_{i+1})+\sigma(\bigcup_{j=1}^l (C_{i+1}\cap D_j))\}dG.
\end{eqnarray}
Because of the definition of $\sigma(\bigcup_{j=1}^l (C_i\cap D_j))$,
\begin{eqnarray}
\int_{\intersect{C}{G}}\sigma(\bigcup_{j=1}^l (C_i\cap D_j))dG
&=&\int_{\intersect{C}{G}}\{\sum_{j=1}^l \sigma(C_i\cap D_j)-\sum_{1\leq j_1<j_2\leq l}\sigma(C_i\cap D_{j_1}\cap D_{j_2})+\cdots \}dG.
\end{eqnarray}
Note that $C_i\cap D_{j_1}\cap D_{j_2}\cap \cdots$ is convex.
Due to a similar equation to Eq. (\ref{sigma_0}), we obtain the following equation.
\begin{eqnarray}
&&\int_{\intersect{C}{G}, \{\intersect{C}{D_i}\}_{i=1}^l}\sigma_i dG\  (dD)^l\cr
&=&\pi\int_{\{\intersect{C}{D_i}\}_{i=1}^l}\{\sizex{C_i}-\sum_{j=1}^l \sizex{C_i\cap D_j}+\sum_{1\leq j_1<j_2\leq l}\sizex{C_i\cap D_{j_1}\cap D_{j_2}}-\cdots \cr
&&-\sizex{C_{i+1}}+\sum_{j=1}^l \sizex{C_{i+1}\cap D_j}-\sum_{1\leq j_1<j_2\leq l}\sizex{C_{i+1}\cap D_{j_1}\cap D_{j_2}}+\cdots\}(dD)^l.
\end{eqnarray}
Because the denominator of Eq. (\ref{chord0}) is $\lengthx{C}\int_{\{\intersect{C}{D_i}\}_{i=1}^l}(dD)^l$ and because of Eq. (\ref{size-def}), we obtain Eq. (\ref{sigma_i0}).
$\square$

\subsubsection{Derivation of $T_d$}
Note that $T_d=E[s_w\sigma_w+\sum_{i=1}^n s_i\sigma_i|\intersect{C}{G}, \{\intersect{C}{D_j}\}_{j=1}^l]$.
Because of Eqs. (\ref{sigma_w0}) and (\ref{sigma_i0}),
\begin{eqnarray}
T_d&=&\frac{\pi}{\lengthx{C}} \sum_{m=0}^l (-1)^{m+1}\sum_{1\leq i_1<\cdots<i_m\leq l} \cr
&&E[s_w\sizex{C\cap_{k=i_1,\cdots,i_m}D_k}\bfone(m>0)-\sum_{i=1}^ns_i(\sizex{C_i\cap_{k=i_1,\cdots,i_m}D_k}-\sizex{C_{i+1}\cap_{k=i_1,\cdots,i_m}D_k})\ |\{\intersect{C}{D_j}\}_{j=1}^l]\qquad.
\end{eqnarray}
By replacing $C_j$ with $C$ in Result \ref{size-result} (that is, Eqs. (\ref{homo-1}), (\ref{pj1}), (\ref{homo-0}), and (\ref{g2})),  $E[\sizex{C\cap_{k=i_1,\cdots,i_m}D_k}\ |\{\intersect{C}{D_j}\}_{j=1}^l]=\sizex{C}\Pi_{k=i_1,\cdots,i_m}(2\pi\sizex{D_k}/f(C,D_k))$ for a homogeneous Poisson process and $E[\sizex{C\cap_{k=i_1,\cdots,i_m}D_k}\ |\{\intersect{C}{D_j}\}_{j=1}^l]\approx g_2(C,m)\Pi_{k=i_1,\cdots,i_m}(2\pi\sizex{D_k}/g_1(C,D_k))$ for an inhomogeneous Poisson process.
In addition, by using Eqs. (\ref{homo-1}) and (\ref{homo-0}) for a homogeneous Poisson process and by using Eqs. (\ref{g2}) and (\ref{pj1}), we obtain the following result.
\begin{result}
For an inhomogeneous Poisson process,
\begin{eqnarray}
T_d&\approx&\frac{\pi}{\lengthx{C}}\sum_{m=0}^l(-1)^{m+1}\{s_wg_2(C,m)\bfone(m>0) -\sum_{i=1}^n s_i(g_2(C_i,m)-g_2(C_{i+1},m))\}b_h(m|C),\label{td-inhomo}
\end{eqnarray}
and for a homogeneous Poisson process,
\begin{eqnarray}
T_d&=&\frac{\pi}{\lengthx{C}}\{s_w\sizex{C}(1-\prod_{k=1}^l(1-2\pi\sizex{D_k}/f(C,D_k))) +\sum_{i=1}^n s_i(\sizex{C_i}-\sizex{C_{i+1}})\prod_{k=1}^l(1-2\pi\sizex{D_k}/f(C,D_k))\}.\label{td-homo}
\end{eqnarray}
\end{result}

\subsubsection{Derivation of $B_d$ from $T_d$ and $q_d(x)$}
Now we are in a position to obtain $B_d$ and $q_d(x)$.
\begin{result}
Under an inhomogeneous Poisson process, $B_d$ and $q_d(x)$ are approximately given by
\begin{eqnarray}
B_d&\approx&\sum_{m=0}^l(-1)^{m+1}\{s_wg_2(C,m)\bfone(m>0) -\sum_{i=1}^n s_i(g_2(C_i,m)-g_2(C_{i+1},m))\}b_h(m|C)/\sizex{C}\label{b_d_i}\cr
&&\\
q_d(x)&\approx&\frac{1}{\sizex{C}}\sum_{m=0}^l(-1)^{m+1}\{\bfone(s_w\geq x,m>0)g_2(C,m) \cr
&&\qquad\quad \qquad\quad -\sum_{i=1}^n\bfone(s_i\geq x)(g_2(C_i,m)-g_2(C_{i+1},m))\} b_h(m|C).\label{q_d_i}
\end{eqnarray}

Under a homogeneous Poisson process, these approximation formulas become exact and simplified into 
\begin{eqnarray}
B_d&=&s_w-\sum_{i=1}^n (s_w-s_i)\frac{\sizex{C_i}-\sizex{C_{i+1}}}{\sizex{C}}\Pi_{j=1}^l  (1-\frac{2\pi\sizex{D_{j}}}{f(C,D_j)}).\label{b_d_h}\\
q_d(x)&=&\sum_{j=1}^n \frac{\sizex{C_{j}}-\sizex{C_{j+1}}}{\sizex{C}}[\bfone(s_w\geq x)\{1-\Pi_{i=1}^l(1-\frac{2\pi\sizex{D_i}}{f(C,D_i)})\}+\bfone(s_j\geq x)\Pi_{i=1}^l(1-\frac{2\pi\sizex{D_i}}{f(C,D_i)})].\label{q_d_h}
\end{eqnarray}
\end{result}

\proof
Because $B_=T_d\lengthx{C}/(\pi\sizex{C})$, Eq. (\ref{b_d_i}) is derived directly from Eq. (\ref{td-inhomo}).
In addition, because of $\sizex{C}=\sum_{j=1}^n \sizex{C_{j}}-\sizex{C_{j+1}}$, Eq. (\ref{b_d_h}) is derived from Eq. (\ref{td-homo}).

Because of the definition of $q_d(x)$ given by Eq. (\ref{def_qd}),
\begin{eqnarray}
q_d(x)=\frac{\int_{\intersect{C}{G}, \{\intersect{C}{D_i}\}_{i=1}^l}dG\  (dD)^l}{\int_{\intersect{C}{G}, \{\intersect{C}{D_i}\}_{i=1}^l}\sigma(C)dG\  (dD)^l}
E[\bfone(s_w\geq x) \sigma_w+\sum_{i=1}^n \bfone(s_i\geq x)\sigma_i|\intersect{C}{G}, \{\intersect{C}{D_i}\}_{i=1}^l].
\end{eqnarray}
We can evaluate $E[\bfone(s_w\geq x) \sigma_w+\sum_{i=1}^n \bfone(s_i\geq x)\sigma_i|\intersect{C}{G}, \{\intersect{C}{D_i}\}_{i=1}^l]$ by replacing $s_w$ with $\bfone(s_w\geq x)$ and $s_i$ with $\bfone(s_i\geq x)$ in derivation of Eqs. (\ref{td-inhomo}) and (\ref{td-homo}).
Because $\frac{\int_{\intersect{C}{G}, \{\intersect{C}{D_i}\}_{i=1}^l}dG\  (dD)^l}{\int_{\intersect{C}{G}, \{\intersect{C}{D_i}\}_{i=1}^l}\sigma(C)dG\  (dD)^l}=\lengthx{C}/(\pi\sizex{C})$, we obtain Eq. (\ref{q_d_i}) and (\ref{q_d_h}).
$\square$

Under a homogeneous Poisson process, Eqs. (\ref{b_d_i}) and (\ref{q_d_i}) are identical to Eqs. (\ref{b_d_h}) and (\ref{q_d_h}), respectively.
That is, the approximation formulas Eqs. (\ref{b_d_i}) and (\ref{q_d_i}) become exact under a homogeneous Poisson process.

Due to Eqs. (\ref{homoBs}), (\ref{homoqs}), (\ref{b_d_h}), and (\ref{q_d_h}), 
\begin{result}
$B_s= B_d$ and $q_s(x)=q_d(x)$ under a homogeneous Poisson process.
\end{result}

In addition, we can see that Eqs. (\ref{inhomoBs}), (\ref{inhomoqs}) are identical to Eqs. (\ref{b_d_i}) and (\ref{q_d_i}) when $\sizex{D_i}=\sizex{D}$ and $\lengthx{D_i}=\lengthx{D}$ for all $i$.
That is, the approximation formulas for $B_s$ $(q_s(x))$ are also applicable as those for $B_d$ $(q_d(x))$ when $\sizex{D_i}=\sizex{D}$ and $\lengthx{D_i}=\lengthx{D}$, even under an inhomogeneous Poisson process.

Therefore, it is strongly suggested that the bit rate a user experiences does not depend on user mobility.

\subsection{Derivation of $N_h$}
Vertical handovers are handovers between a cellular network and a WLAN, and occur when a user moves from one network to the other network.
Generally, vertical handovers occur at the boundary of $C$ or that of $D_i$ ($i=1,2,\cdots$). We focus on the vertical handovers at the boundary of $D_i$ within $C$ occurring when a user moving along $G$ passes through the boundary of $D_i$ from outside $D_i$ into $D_i$ within $C$ or from inside $D_i$ to outside $D_i$ within $C$ if the passing point $\partial D_i\cap G$ is not covered by any other WLANs.

\begin{figure}[htb] 
\begin{center} 
\includegraphics[width=6cm,clip]{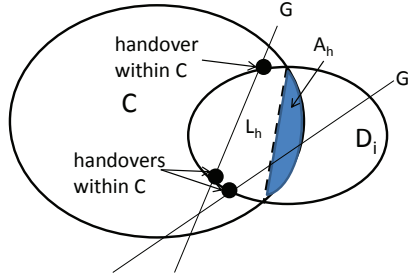} 
\caption{$A_h$ and $L_h$} 
\label{A_h} 
\end{center} 
\end{figure}

To evaluate $N_h$, we define $L_h$ and $A_h$.
When $\intersect{D_i}{\partial C}$, we can define $L_h$ as a line segment between two end points of $D_i\cap\partial C$ and $A_h$ as a convex region surrounded by $L_h$ and $D_i\cap\partial C$ in $D_i$ (Fig. \ref{A_h}).

Let $S_i$ be the number of intersections between $G$ and $\partial {D_i}$, $E[S_i\vert {\rm E1}]$ be its expectation when $D_i\subset C,\intersect {C}{G}$, and $E[S_i\vert {\rm E2}]$ be its expectation when $\intersect{D_i}{\partial C},\intersect {C}{G}$.
Because $\intersect {C}{G}$ is satisfied when $\intersect{G}{D_i}, D_i\subset C$ or when $\intersect {G}{L_h}$,
\begin{eqnarray}
E[S_i\vert {\rm E1}]&=&2\Pr(\intersect{G}{D_i}, D_i\subset C)\cr
&=&\frac{2\int_{D_i\subset C,\intersect {D_i}{G}}dG\, dD_i}{\int_{\intersect {C}{G}, \intersect{C}{D_i}}dG \  dD_i},\label{e1}
\end{eqnarray}
According to Eq. (\ref{3.12}) and because of $A_h\subset D_i\cap C$,
\begin{eqnarray}
E[S_i\vert {\rm E2}]&=&2\Pr(G\cap A_h=\emptyset,\intersect {D_i\cap C}{G},\intersect{D_i}{\partial C})+\Pr(\intersect {G}{L_h},\intersect{D_i}{\partial C})\cr
&=&\frac{\int_{\intersect{D_i}{\partial C}}\{2\int_{\intersect {D_i\cap C}{G}} dG-2\int_{\intersect{G}{A_h}} dG+\int_{\intersect {G}{L_h}}dG\}dD_i}{\int_{\intersect {C}{G}, \intersect{C}{D_i}}dG \  dD_i}\cr
&=&\frac{\int_{\intersect{D_i}{\partial C}}(2\lengthx{D_i\cap C}-2\lengthx{A_h}+\lengthx{L_h})dD_i}{\int_{\intersect {C}{G}, \intersect{C}{D_i}}dG \  dD_i}\cr
&=&\frac{\int_{\intersect{D_i}{\partial C}}2\arc{D_i}{C}dD_i}{\int_{\intersect {C}{G}, \intersect{C}{D_i}}dG \  dD_i}.\label{e2}
\end{eqnarray}

Because $D_j$ ($j\neq i$) is independent of $G$ and $D_i$, 
\begin{eqnarray}
N_h&=&\sum_{i=1}^l(E[S_i\vert {\rm E1}]+E[S_i\vert {\rm E2}])\Pr((G\cap\partial {D_i})\not\in \bigcup_{j\neq i} D_j|\{\intersect{C}{D_j}\}_{j\neq i})\label{e12}
\end{eqnarray}
where $\Pr((G\cap\partial {D_i})\not\in \bigcup_{j\neq i} D_j|\{\intersect{C}{D_j}\}_{j\neq i})$ is a conditional probability that an intersection between $G$ and $\partial {D_i}$ is not covered by any of $D_j$ $j\neq i$ with the condition that $\{\intersect{C}{D_j}\}_{j\neq i}$.

Then, we can derive the following result.
\begin{result}
For a homogeneous Poisson process,
\begin{eqnarray}
N_h&=&\sum_{i=1}^l  \frac{4\pi\sizex{C}\cdot\lengthx{D_i}}{\lengthx{C}f(C,D_i)}\Pr((G\cap\partial {D_i})\not\in \bigcup_{j\neq i} D_j|\{\intersect{C}{D_j}\}_{j\neq i}).\label{nh_homo}
\end{eqnarray}
For an inhomogeneous Poisson process
\begin{eqnarray}
N_h&\approx&\sum_{i=1}^l  \frac{4\pi(\sizex{C}+\rho_H\sizex{C\cap\Omega_H}-\rho_L\sizex{C\cap\Omega_L})}{g_1(C,D_i)\lengthx{C}} \lengthx{D_i}\Pr((G\cap\partial {D_i})\not\in \bigcup_{j\neq i} D_j|\{\intersect{C}{D_j}\}_{j\neq i}).\label{nh_inhomo}
\end{eqnarray}
\end{result}

\noindent {\bf Proof of Eq. (\ref{nh_homo}):}
Assume a  homogeneous Poisson process.
Due to Eqs. (\ref{3.12}) and (\ref{6.52}),
\begin{eqnarray}
&&\int_{D_i\subset C,\intersect {D_i}{G}}dG\, dD_i=\int_{D_i\subset C}\lengthx{D_i} dD_i=\lengthx{D_i} \{2\pi (\sizex{D_i}+\sizex{C})-\lengthx{D_i}\cdot\lengthx{C}\},
\end{eqnarray}
and due to Eqs. (\ref{3.12}) and (\ref{6.48}),
\begin{eqnarray}
&&\int_{\intersect {C}{G}, \intersect{C}{D_i}}dG \  dD_i=\lengthx{C}f(C,D_i).
\end{eqnarray}
Apply these two equations to Eq. (\ref{e1}).

Due to Eq. (\ref{6.61}), 
\begin{eqnarray}
\int_{\intersect{D_i}{\partial C}}\arc{D_i}{C}dD_i=\lengthx{D_i}(\lengthx{C}\cdot\lengthx{D_i}-2\pi\sizex{D_i}).
\end{eqnarray}
Apply this equation to Eq. (\ref{e2}).
Therefore, we obtain Eq. (\ref{nh_homo}).  $\square$

\noindent {\bf Proof of Eq. (\ref{nh_inhomo}):}
Assume an inhomogeneous Poisson process.
According to Eqs. (\ref{3.12}), (\ref{6.52}), (\ref{6.52r}), and (\ref{inhomo0}) with $\phi=1, X_c=D_i\subset C$,
\begin{eqnarray}
&&\int_{\intersect {D_i}{G},D_i\subset C}dG\,dD_i\cr
&\approx&\rho_0\lengthx{D_i} \{\int_{D_i\subset C,\bfw_i\in {\mathbb R}^2} d\bfw_i\  d\gamma_i+\rho_H\int_{D_i\subset C,\bfw_i\in \Omega_H} d\bfw_i\  d\gamma_i-\rho_L\int_{D_i\subset C,\bfw_i\in \Omega_L}d\bfw_i\  d\gamma_i\}\cr
&\approx&\rho_0g_3(C,D_i)
\end{eqnarray}
where $g_3(C,D_i)\defeq \lengthx{D_i}\{2\pi (\sizex{D_i}+\sizex{C})-\lengthx{D_i}\cdot\lengthx{C}+2\pi\rho_H (\eqSecond{D_i}{C}{\Omega_H})-2\pi\rho_L(\eqSecond{D_i}{C}{\Omega_L})\}$.
Apply this equation and Eqs. (\ref{3.12}) and (\ref{dDi}) to Eq. (\ref{e1}).

Due to Eqs. (\ref{6.61}), (\ref{6.61r}), and (\ref{inhomo0}) with $\phi=\arc{D_i}{C}, X_c=\intersect{D_i}{\partial C}$,
\begin{eqnarray}
&&\int_{\intersect{D_i}{\partial C}}\arc{D_i}{C}dD_i\cr
&=&\rho_0\{\int_{\intersect{D_i}{\partial C},\bfw_i\in {\mathbb R}^2}\arc{D_i}{C}d\bfw_i\  d\gamma_i+\rho_H\int_{\intersect{\partial C}{D_i},\bfw_i \in \Omega_H}\arc{D_i}{C}d\bfw_i\  d\gamma_i\cr
&&\qquad - \rho_L\int_{\intersect{\partial C}{D_i},\bfw_i \in \Omega_L}\arc{D_i}{C}d\bfw_i\  d\gamma_i\}\cr
&\approx& \rho_0\lengthx{D_i}\{(\lengthx{C}\cdot\lengthx{D_i}-2\pi \sizex{D_i})+2 \pi \rho_H\hat r(D_i)\arc{C}{\Omega_H}-2 \pi \rho_L\hat r(D_i)\arc{C}{\Omega_L}\}.
\end{eqnarray}
Apply this equation and Eqs. (\ref{3.12}) and (\ref{dDi}) to Eq. (\ref{e2}).
As a result, we obtain Eq. (\ref{nh_inhomo}).
$\square$

Now, we need to evaluate $\Pr((G\cap\partial {D_i})\not\in \bigcup_{j\neq i} D_j|\{\intersect{C}{D_j}\}_{j\neq i})=\frac{\int_{(G\cap\partial {D_i})\not\in \bigcup_{j\neq i} D_j, \{\intersect{C}{D_j}\}_{j\neq i}}((dD)^l/dD_i)}{\int_{ \{\intersect{C}{D_j}\}_{j\neq i}}((dD)^l/dD_i)}$ where where $((dD)^l/dD_i)\defeq dD_1\cdots dD_{i-1}\,dD_{i+1}\cdots dD_l$.

\begin{result}
For a homogeneous Poisson process,
\begin{eqnarray}
\Pr((G\cap\partial {D_i})\not\in \bigcup_{j\neq i} D_j|\{\intersect{C}{D_j}\}_{j\neq i})=\Pi_{j\neq i}(1-2\pi\sizex{D_j}/f(C,D_j)).
\end{eqnarray}
For an inhomogeneous Poisson process,
\begin{eqnarray}
\Pr((G\cap\partial {D_i})\not\in \bigcup_{j\neq i} D_j|\{\intersect{C}{D_j}\}_{j\neq i})\approx\Pi_{j\neq i}(1-2\pi\sizex{D_j}g_4(C)/g_1(C,D_j))
\end{eqnarray}
where $g_4(C)\defeq 1+(\rho_H\lambda_H\sizex{C\cap\Omega_H}-\rho_L\lambda_L\sizex{C\cap\Omega_L})/(\lambda_C\sizex{C})$.
\end{result}

\proof
Note that $\int_{(G\cap\partial {D_i})\not\in D_j, \intersect{C}{D_j}}dD_j=\int_{(G\cap\partial {D_i})\not\in D_j, \intersect{C}{D_j}}dD_j=\int_{\intersect{C}{D_j}}dD_j-\int_{(G\cap\partial {D_i})\in D_j}dD_j$, because $(G\cap\partial {D_i})\subset C$.

Therefore, for a homogeneous Poisson process, due to Eqs. (\ref{6.48}) and (\ref{point}), for $j\neq i$,
\begin{equation}
\int_{(G\cap\partial {D_i})\not\in D_j, \intersect{C}{D_j}}dD_j=f(C,D_j)-2\pi\sizex{D_j},\label{not1}
\end{equation}
By repeatedly using Eqs. (\ref{6.48}) and (\ref{not1}),
\begin{eqnarray}
&&\Pr((G\cap\partial {D_i})\not\in \bigcup_{j\neq i} D_j|\{\intersect{C}{D_j}\}_{j\neq i})=\Pi_{j\neq i}(1-2\pi\sizex{D_j}/f(C,D_j)).
\end{eqnarray}

For an inhomogeneous Poisson process, due to Eqs. (\ref{inhomo0}) with $\phi=1, X_c=(G\cap\partial {D_i})\in D_j$ and (\ref{dDi}),
\begin{eqnarray}
&&\int_{(G\cap\partial {D_i})\not\in D_j, \intersect{D_j}{C}}dD_j\cr
&\approx&\rho_0g_1(C,D_j)\cr
&&\qquad-\rho_0\{\int_{(G\cap\partial {D_i})\in D_j,\bfw_i\in {\mathbb R}^2} d\bfw_i\  d\gamma_i+\rho_H\int_{(G\cap\partial {D_i})\in D_j,\bfw_i\in \Omega_H} d\bfw_i\  d\gamma_i-\rho_L\int_{(G\cap\partial {D_i})\in D_j,\bfw_i\in \Omega_L}d\bfw_i\  d\gamma_i\}\cr
&\approx&\rho_0g_1(C,D_j)-2\pi\rho_0\sizex{D_j}\{1+\rho_H\Pr((G\cap\partial {D_i})\in \Omega_H)-\rho_L\Pr((G\cap\partial {D_i})\in \Omega_L)\}.
\end{eqnarray}
The last equality above is due to Eqs. (\ref{point}) and (\ref{6.48rr}).

It is reasonable to assume that $\Pr((G\cap\partial {D_i})\in \Omega_H)$ ($\Pr((G\cap\partial {D_i})\in \Omega_L)$) is proportional to the ratio of the mean number of WLANs in $\Omega_H\cap C$ ($\Omega_L\cap C$) to that in $C$.
That is, assume that $\Pr((G\cap\partial {D_i})\in \Omega_H)\approx\frac{\lambda_H\sizex{C\cap\Omega_H}}{\lambda_H\sizex{C\cap\Omega_H}+\lambda_L\sizex{C\cap\Omega_L}+\lambda_0(\sizex{C}-\sizex{C\cap\Omega_H}-\sizex{C\cap\Omega_L})}=\frac{(1+\rho_H)\sizex{C\cap\Omega_H}}{\sizex{C}+\rho_H\sizex{C\cap\Omega_H}-\rho_L\sizex{C\cap\Omega_L}}=\lambda_H\sizex{C\cap\Omega_H}/(\lambda_C\sizex{C})$, and $\Pr((G\cap\partial {D_i})\in \Omega_L)\approx\lambda_L\sizex{C\cap\Omega_L}/(\lambda_C\sizex{C})$.
Therefore,
\begin{eqnarray}
\int_{(G\cap\partial {D_i})\not\in D_j, \intersect{D_j}{C}}dD_j
&\approx&\rho_0(g_1(C,D_j)-2\pi \sizex{D_j}g_4(C)).
\end{eqnarray}
By repeatedly applying this equation and Eq. (\ref{dDi}),
\begin{eqnarray}
&&\Pr((G\cap\partial {D_i})\not\in \bigcup_{j\neq i} D_j|\{\intersect{C}{D_j}\}_{j\neq i})\approx\Pi_{j\neq i}(1-2\pi\sizex{D_j}g_4(C)/g_1(C,D_j)).
\end{eqnarray}

Consequently, we obtain the following result.
\begin{result}
The mean number $N_h$ of vertical handovers is approximately given by the following formula under an inhomogeneous Poisson process.
\begin{eqnarray}
N_h&\approx&\frac{4\pi\sum_{i=1}^l \lengthx{D_i}(\sizex{C}+\rho_H\sizex{C\cap\Omega_H}-\rho_L\sizex{C\cap\Omega_L})\Pi_{j\neq i}(g_1(C,D_j)-2\pi\sizex{D_j}g_4(C))}{\lengthx{C}\Pi_{i=1}^l g_1(C,D_i)}\label{n_v}
\end{eqnarray}

Under a homogeneous Poisson process, this approximation formula becomes exact and simplified into
\begin{eqnarray}
N_h&=&\frac{4\pi\sizex{C}\sum_{i=1}^l \lengthx{D_i}\Pi_{j\neq i}(f(C,D_j)-2\pi\sizex{D_j})}{\lengthx{C}\Pi_{i=1}^l f(C,D_i)}.\label{n_v_h}
\end{eqnarray}

\end{result}

\section{Numerical examples}
\subsection{Conditions of numerical examples}
\begin{figure}[tb] 
\begin{center} 
\includegraphics[width=6cm,clip]{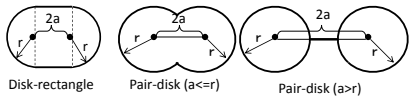} 
\caption{Shape of $D$} 
\label{WiFi} 
\end{center} 
\end{figure}

In the numerical examples in this section, we assume that all $D_i$ are
congruent for any $i$ and are disk-rectangles or pair-disks, as shown in
Fig.~\ref{WiFi}, with $r=50$ m (disk-shaped $D$ means $a=0$). We set
$r=50$ m based on a measurement study~\cite{chen01} or
information~\cite{wifi_wiki} where the range of WiFi APs is reported as dozens of meters to over $100$ m.  

To determine the parameter setting of bitrates $s_i$ for cellular networks and $s_w$ for
WLANs, we use the results of measurement studies in actually operated
WiFi and 3G networks~\cite{deshpande-IMC2010,balasubramanian-MobiSys2010,gass-pam2010}. 
They measured throughput from a server with a public IP address to mobile
terminal(s) with a WiFi interface and that with a 3G interface under various
access scenarios such as walking or driving for a long/short distance. Deshpande et al.~\cite{deshpande-IMC2010} used a
commercially operated metro-scale WiFi network while Balasubramanian et al.~\cite{balasubramanian-MobiSys2010} and Gass and Diot~\cite{gass-pam2010} used open
WiFi APs in the wild. 
In~\cite{deshpande-IMC2010,balasubramanian-MobiSys2010,gass-pam2010}, the measured throughput of the WLAN ranged from $1-2$ (Mbps) at
median or average to around $5-10$ (Mbps) at maximum. For 3G networks, the throughput ranged
from about $500$ (kbps) at median or average to around $1.5-2$ (Mbps) at
maximum. Another performance metric is availability. Balasubramanian et al.~\cite{balasubramanian-MobiSys2010} reported that the availability of the 3G network is $87\%$ while that
of the WiFi network is $11\%$. Because of the low availability of WiFi, the median or
average throughput becomes small, so in our numerical example, we use the maximum
throughput of WiFi as the bitrate of a WLAN cell $s_w=10,000$ (kbps) rather
than average or median of throughput, assuming that the offloading is
executed when the terminal moves into a WLAN cell and is expected to
achieve sufficient throughput. 
(The throughput measurement studies of a WLAN network in experimental environments, such
as~\cite{chen01,wijesinha-11g-2005}, show that the maximum throughput is
about $10-20$ (Mbps).)
\OMIT{
In another related work, Bruno et al.~\cite{bruno-MobiCom2008}
investigated throughput performance; however, they focused on the case
in which there exists cross traffic rather than measuring the maximum throughput.)
}

For cellular network parameters, we use the distribution of throughput in~\cite{deshpande-IMC2010} as follows. Assuming that a user
location is uniformly distributed in a cell, we first calculate the
frequency (in percentage) $F_i$ of users existing in $C_i-C_{i+1}$ where
$C_i$ are disk-shaped with radii $1000, 500, 200, 100$ (m) for
$i=1,2,3,4$. For example, $F_4=100^2/1000^2\times 100=1\%$ and
$F_3=(200^2-100^2)/1000^2\times 100=3\%$. We also assume that the
throughput, i.e., the achieved bitrate (bps) $s_i$, in $C_i-C_{i+1}$
is higher than $s_{i-1}$. We set $s_i (i=1,2,3)$ so that the $s_i$ is between the
$100-\sum_{j=i}^4 F_j$ and $100-\sum_{j=i+1}^4 F_j$ percentile of
throughput, i.e., $s_i$ is set to the $100-(\sum_{j=i}^4 F_j+\sum_{j=i+1}^4 F_j)/2$
percentile. For $s_4$, we set $s_4$ to be the $100-F_4/2$ percentile of throughput.
Based on the graph in~\cite{deshpande-IMC2010}, we set $s_1=300$, $s_2=750$,
$s_3=1500$, and $s_4=2000$ (kbps). 

\OMIT{
Note that there have been other studies related to 3G network
measurements~\cite{shafiq-sigmetrics2011,shafiq-sigmetrics2012,gerber-IMC2010}. However,
Shafiq et al.~\cite{shafiq-sigmetrics2011,shafiq-sigmetrics2012} focused on
traffic characteristics, such as time-series in traffic volume, rather
than throughput performance. Gerber et al.~\cite{gerber-IMC2010} proposed a method of estimating achievable download
throughput from passive measurement and analyzed the throughput
characteristics in 3G networks. However, the throughput value is
normalized by some constant due to confidentiality.
}

\subsection{Accuracy of proposed formulas}
We compare the values of the performance metrics derived by the derived formulas and those obtained by simulation, where the empirical data of the WLAN APs of three network operators are used.
We provide the center of $C$ at one of the nine 1200 $\times$ 1200-m square grids in which the upper left corner is located at (1200 m, -1200 m) on the map in Fig. \ref{map}.
For this location of $C$, we conducted simulation.
Because we change the location of $C$ to another one of the nine grids, we can obtain nine simulation results for each operator.

To evaluate the metric theoretically, $\Omega_H$ and $\Omega_L$ used here are identified using the following method.
(1) Divide the 5 $\times$ 5-km square region into 100 $\times$ 100-m subregions called atoms, and set $\omega_H=\omega_L=\emptyset$.
(2) Count the number of WLAN APs in each atom. Let $a_0$ be the average number of WLAN APs in an atom.
(3) Consider a window defined by consecutive $n_0 \times n_0$ atoms. Let $n_b$ be the number of WLAN APs in a window and $a_U$ and $a_L$ be the 0.999 upper/lower quantile of $n_b$ under a homogeneous Poisson process: $a_U\defeq a_0n_0^2+3\sqrt{a_0}n_0$ and $a_L\defeq a_0n_0^2-3\sqrt{a_0}n_0$. 
If $n_b>a_U$ ($n_b<a_L$), atoms in the window are determined as atoms in $\omega_H$ ($\omega_L$).
(4) Slide the window and repeat (3) until the window sweeps the entire 5 $\times$ 5-km square region.
(5) Atoms belonging to $\omega_H$ ($\omega_L$) more than $n_0^2/2$ times are determined as those belonging to $\Omega_H$ ($\Omega_L$).
Note that $\Omega_H$ ($\Omega_L$) may not be consecutive.

In this paper, the set of WLAN AP data of operator (c) is used to identify $\Omega_H$ and $\Omega_L$ because the number of APs is much larger than those of the other operators.
We used $n_0=3$.
As a result, we obtained $\sizex{\Omega_H}=3.48$ km$^2$, $\sizex{\Omega_L}=5.31$ km$^2$; $\lambda_0$=23.87, $\lambda_H$=27.30,  $\lambda_L$=4.33, $\rho_H$=0.143, and $\rho_L$=0.819, for operator (a); $\lambda_0$=5.305, $\lambda_H$=20.40,  $\lambda_L$=1.318, $\rho_H$=2.846, and $\rho_L$=0.752, for operator (b); and $\lambda_0$=103.8, $\lambda_H$=575,  $\lambda_L$=3.766, $\rho_H$=4.538, and $\rho_L$=0.964, for operator (c).
In addition to these parameter values, we obtained the number $l$ of WLANs intersecting $C$, $\sizex{C_j \cap \Omega_H}$, $\sizex{C_j \cap \Omega_L}$, $\arc{C}{\Omega_H}$, and $\arc{C}{\Omega_L}$ in the simulation, and used them in the theoretical evaluation.

\begin{figure}[tb] 
\begin{center} 
\includegraphics[width=6cm,clip]{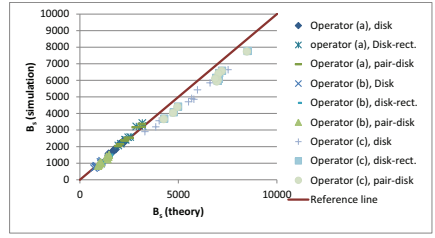} 
\caption{Comparison between theory and simulation for $B_s$} 
\label{B_sAccuracy} 
\end{center} 
\end{figure}

Figure \ref{B_sAccuracy} shows the comparison results for $B_s$.
Each point corresponds to each location of $C$.
Although there is a small bias for operator (c), the agreement between the theoretical and simulation results is, in general, very good.
Therefore, we conclude that the derived approximation formula works fine.

\begin{figure}[tb] 
\begin{center} 
\includegraphics[width=6cm,clip]{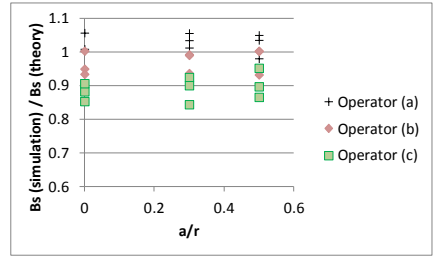} 
\caption{Ratio of $B_s$ derived by simulation to that derived by theory for pair-disk shaped $C_i$} 
\label{non-convex} 
\end{center} 
\end{figure}

We also investigated the accuracy of the proposed model for a non-convex $C_i$.
$C_i$ is assumed to be pair-disk shaped (Fig.~\ref{WiFi}) for all $i$ and the middle points of two disk centers for all $i$ are located at the same point.
The line between two disk centers of $C_i$ is parallel to the $x$-axis in Fig. \ref{map}.
Figure \ref{non-convex} plots the ratio of $B_s$ derived by simulation to that derived by theory where the $x$-axis denotes the ratio $a/r$ in Fig. \ref{WiFi} and each point in this figure corresponds to the locations of the middle point of two disk centers of $C_i$.
This figure shows that there is no clear accuracy deterioration, although a pair-disk with $a/r>0$ is not convex.

We also simulated $B_d$ by drawing 1000 lines as $G$ for each case and compared 27 cases of $B_s$ and $B_d$ (nine locations of the center of $C$ and three shapes for $D$) for each operator.
The relative absolute difference $|B_d-B_s|/B_d$ was at most 3.23\%, 3.38\%, and 2.48\%, and the average was 1.51\%, 1.59\%, and 0.640\% for operators (a), (b), and (c).
Therefore, we can say that $B_d=B_s$ is valid.
As a result, our derived approximation formula also works fine for $B_d$ because it works fine for $B_s$.

\begin{figure}[tb] 
\begin{center} 
\includegraphics[width=6cm,clip]{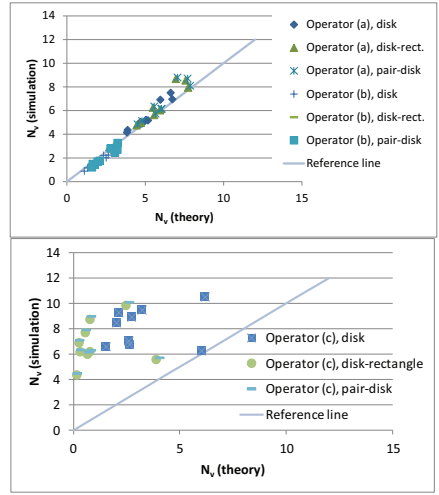} 
\caption{Comparison between theory and simulation for $N_v$} 
\label{N_vAccuracy} 
\end{center} 
\end{figure}

Figure \ref{N_vAccuracy} compares $N_v$ derived using the derived formula and the simulation.
For operators (a) and (b), the theoretical and simulation results agree well.
On the contrary, the derived formula largely underestimates $N_v$ in most cases for operator (c).
This is because $l$ (the number of WLANs intersecting $C$) of operator (c) is much larger than 100 and is sometimes larger than 1000, and because Eq. (\ref{n_v}) includes the product term $(1-\frac{2\pi \sizex{D_j}g_4(C)}{g_1(C,D_j)})^l$.
When there is a small approximation error or modeling error in $1-\frac{2\pi \sizex{D_j}g_4(C)}{g_1(C,D_j)}$, the error becomes enormous in $N_v$.
For example, when $l=1000$, $0.99^l/0.98^l\approx 2.5\times10^4$.
That is, the difference of 1\% results in a difference of four orders of magnitude.
Because $l$ for operator (a) is nearly 100, and that for operator (b) is much less than 100, we need to be careful when using the derived formula for $l\gg 100$.

\subsection{Evaluation of metrics}
In the remainder of this paper, the following are the parameter values we use in the evaluation in addition to those described at the beginning of this section, unless explicitly indicated otherwise.
These parameter values are called the default values:
$\sizex{C\cap\Omega_H}=\sizex{C\cap\Omega_L}=0.3\sizex{C}$, $a=0$ (that is, $D$ is disk-shaped), $\rho_H=3$, $\rho_L=1$, $\arc{C}{\Omega_H}=2\sqrt{2\sizex{C\cap\Omega_H}/\pi}$, and $\arc{C}{\Omega_L}=2\sqrt{2\sizex{C\cap\Omega_L}/\pi}$.
The default values of $\arc{C}{\Omega_H}$ and $\arc{C}{\Omega_L}$ are based on the assumption that $C\cap\Omega_H$ ($C\cap\Omega_L$) is a semicircle and $\arc{C}{\Omega_H}$ ( $\arc{C}{\Omega_L}$) is its diameter.

\begin{figure}[tb] 
\begin{center} 
\includegraphics[width=6cm,clip]{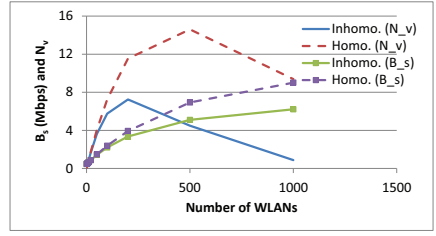} 
\caption{Comparison between inhomogeneous Poisson process and homogeneous Poisson process for $B_s$ and $N_v$} 
\label{Poisson} 
\end{center} 
\end{figure}

First, we compare $B_s(=B_d)$ and $N_v$ under the inhomogeneous Poisson process with the default parameter values and those under the homogeneous Poisson process.
The results are plotted in Fig. \ref{Poisson}.
As $l$ (the number of WLANs) increases, the difference in $B_s$ under these processes becomes quite large.
When $l=1000$, $B_s$ under the homogeneous Poisson process is overestimated by about 50\% compared to $B_s$ under the inhomogeneous Poisson process.
Furthermore, (1) $N_v$ is not monotone against $l$, and (2) $N_v$ under the homogeneous Poisson process is extremly overestimated compared to $N_v$ under the inhomogeneous Poisson process.
Fact (2) is natural because $\{D_i\}_i$ under the inhomogeneous Poisson process is more likely to clump than that under the homogeneous Poisson process.
Therefore, $\{D_i\}_i$ is likely to make a cluster, and the vertical handover does not occur in a cluster of $\{D_i\}_i$.

We also compare $r_{WLAN}$ (the ratio in which the traffic is delivered through the WLANs) and $p_{WLAN}$(the raito of $C$ covered by WLANs) under the inhomogeneous Poisson process with the default parameter values and those under the homogeneous Poisson process.
Assume that traffic is uniformly generated in $C$.
If the point where traffic is generated is covered by a WLAN AP, the traffic is assumed to be delivered through the WLAN with achieved bitrate $s_w$ (bps); otherwise, it is delivered through the cellular network with achieved bitrate $s_1,\cdots, s_n$ (bps).
The ratio $r_{WLAN}$ is approximately given by $s_w(1-\sum_{j=1}^n (b_0g_0(j)+b_Hg_H(j)+b_Lg_L(j))/\sizex{C})/B_s$.
Similarly, the ratio $p_{WLAN}$ is approximately given by $\sum_{j=1}^n p_j \approx 1-\sum_{j=1}^n (b_0g_0(j)+b_Hg_H(j)+b_Lg_L(j))/\sizex{C}$.

\begin{figure}[tb] 
\begin{center} 
\includegraphics[width=6cm,clip]{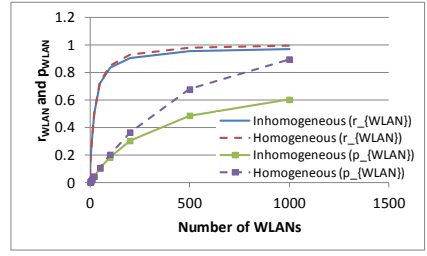} 
\caption{Ratio in which traffic is delivered through WLANs and ratio of $C$ covered by WLANs.}
\label{ratio} 
\end{center} 
\end{figure}

Figure \ref{ratio} shows $r_{WLAN}$ and $p_{WLAN}$.
The former approaches 1 very fast because the difference in $s_w$ and $s_i$ is very large. The difference in ^^ ^^ Inhomogeneous" and ^^ ^^ Homogeneous" is not negligible.
On the other hand, the latter is similar to the curve of $B_s$ in Fig. \ref{Poisson}.
As the number $l$ of WLANs increases, $p_{WLAN}$ becomes larger.
However, its increase becomes smaller as $l$ becomes larger.

\begin{figure}[tb] 
\begin{center} 
\includegraphics[width=6cm,clip]{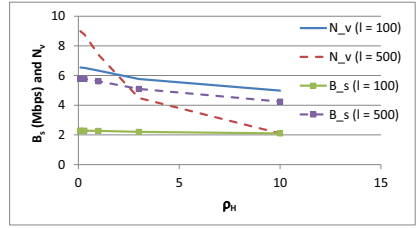} 
\caption{Impact of $\rho_H$ on $B_s$ and  $N_v$} 
\label{rhoH} 
\end{center} 
\end{figure}

\begin{figure}[tb] 
\begin{center} 
\includegraphics[width=6cm,clip]{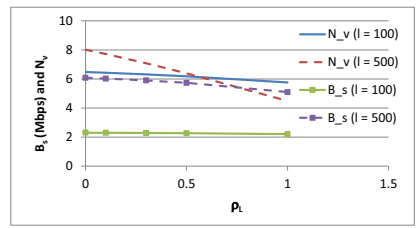} 
\caption{Impact of $\rho_L$ on $B_s$ and  $N_v$} 
\label{rhoL} 
\end{center} 
\end{figure}

Second, we evaluate the impact of $\rho_H$ and $\rho_L$ on $B_s$ and $N_v$ (Figs. \ref{rhoH}, \ref{rhoL}).
When $l=100$, $B_s(=B_d)$ is almost insensitive to $\rho_H$ or $\rho_L$.
When $l=500$, $B_s$ decreases as $\rho_H$ or $\rho_L$ increases.
$N_v$ is also a decreasing function of $\rho_H$ and $\rho_L$.
The sensitivity of $N_v$ to $\rho_H$ or $\rho_L$ becomes larger as $l$ increases. 
Because $\rho_H$ and $\rho_L$ indicate the relative difference in $\lambda_H$ and $\lambda_L$ from $\lambda_0$, this means that $B_s$ and $N_v$ become smaller when inhomogeneity becomes larger.

\begin{figure}[tb] 
\begin{center} 
\includegraphics[width=6cm,clip]{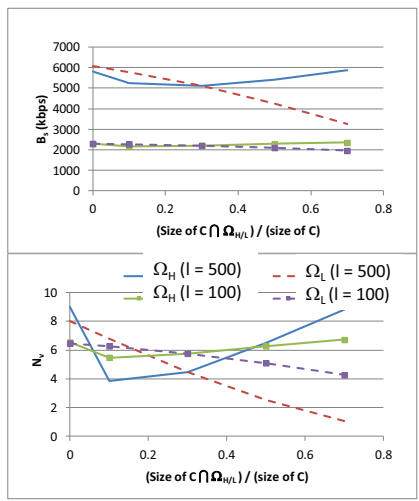} 
\caption{Impact of $\sizex{C\cap\Omega_H}$ and $\sizex{C\cap\Omega_L}$ on $B_s$ and $N_v$} 
\label{omega} 
\end{center} 
\end{figure}

The values $\sizex{C\cap\Omega_H}$ and $\sizex{C\cap\Omega_L}$ are clump parameters of the spatial point process of deployed WLANs.
Figure \ref{omega} plots $B_s(=B_d)$ and $N_v$ as a function of $\sizex{C\cap\Omega_H}/\sizex{C}$ or $\sizex{C\cap\Omega_L}/\sizex{C}$.
Because the default value of $\sizex{C\cap\Omega_H}/\sizex{C}$ ($\sizex{C\cap\Omega_L}/\sizex{C}$) is 0.3, $\sizex{C\cap\Omega_L}/\sizex{C}$ ($\sizex{C\cap\Omega_H}/\sizex{C}$) can move from 0 to 0.7.
In this figure, $B_s$ and $N_v$ are decreasing functions of $\sizex{C\cap\Omega_L}/\sizex{C}$.
This is because when the low WLAN-density region ($\Omega_L$) becomes larger in $C$ with a fixed $l$, many WLAN APs must be in the high or normal WLAN-density region ($\Omega_H$, $\overline{\Omega_H\cup\Omega_L}$), mainly in $\Omega_H$.
As a result, $\{D_i\}_i$ overlap each other in such regions, particularly in $\Omega_H$.
The lower $B_s$ means that $\{D_i\}_i$ cannot achieve efficient coverage due to this overlap, and the smaller $N_v$ is also due to this overlap.
On the other hand, either $B_s$ or $N_v$ is not monotonous against $\sizex{C\cap\Omega_H}/\sizex{C}$ and has the minimum point between $\sizex{C\cap\Omega_H}/\sizex{C}=0$ and 0.7.
This is because (i) when $\sizex{C\cap\Omega_H}/\sizex{C}=0$, WLAN APs are in the normal density region, which occupies 70\% of $C$. Thus, $\{D_i\}_i$ do not overlap so much.  
(ii) When $0<\sizex{C\cap\Omega_H}/\sizex{C}<0.7$, $\{D_i\}_i$ overlap each other in $\Omega_H$ and $\overline{\Omega_H\cup\Omega_L}$, particularly in $\Omega_H$.
Because of the overlaps, efficient coverage cannot be achieved or vertical handover does not occur. (iii) When $\sizex{C\cap\Omega_H}/\sizex{C}=0.7$, WLAN APs are in $\Omega_H\cap C$ but $\{D_i\}_i$ do not overlap so much because $\Omega_H\cap C$ occupies 70\% of $C$.
For $l=500$, $B_s$ is more sensitive to $\sizex{C\cap\Omega_L}/\sizex{C}$ than to $\rho_L$, and more sensitive to $\rho_H$ than to $\sizex{C\cap\Omega_H}/\sizex{C}$.
(Although $\rho_H$ is in [0,10) in Fig. \ref{rhoH}, the range of $\rho_H$ is $[0,\infty)$. Even in $\rho_H\in [0,10)$, the range of $B_s$ shown in Fig. \ref{rhoH} is larger than that of $B_s$ for the full range of $\sizex{C\cap\Omega_H}/\sizex{C}$.)
Therefore, $\sizex{C\cap\Omega_L}/\sizex{C}$ and $\rho_H$ are important parameters for $B_s$.
$N_v$ is also more sensitive to $\sizex{C\cap\Omega_L}/\sizex{C}$ than to $\rho_L$, and to $\rho_H$ than to $\sizex{C\cap\Omega_H}/\sizex{C}$. Thus, $\sizex{C\cap\Omega_L}/\sizex{C}$ and $\rho_H$ are also important parameters for $N_v$.

\begin{figure}[tb] 
\begin{center} 
\includegraphics[width=6cm,clip]{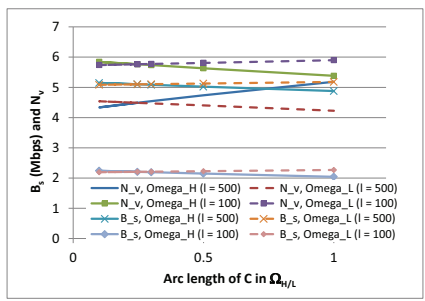} 
\caption{Impact of $\arc{C}{\Omega_H}$ and $\arc{C}{\Omega_L}$ on $B_s$ and $N_v$}
\label{arc} 
\end{center} 
\end{figure}

Figure \ref{arc} plots $B_s(=B_d)$ and $N_v$ as functions of $\arc{C}{\Omega_H}$ and $\arc{C}{\Omega_L}$.
It shows that $B_s$ is a slightly increasing function of $\arc{C}{\Omega_H}$ and a slightly decreasing function of $\arc{C}{\Omega_L}$.
However, $B_s$ is more sensitive to other parameters such as $\sizex{C\cap\Omega_L}/\sizex{C}$ and $\rho_H$.
The behavior of $N_v$ as a function of $\arc{C}{\Omega_H}$ and $\arc{C}{\Omega_L}$ is interesting.
When $l=100$, it is an increasing function of $\arc{C}{\Omega_H}$ and a decreasing function of $\arc{C}{\Omega_L}$.
However, when $l=500$, it is the opposite: it is a decreasing function of $\arc{C}{\Omega_H}$ and an increasing function of $\arc{C}{\Omega_L}$.
As a whole, $N_v$ is more sensitive to other parameters such as $\sizex{C\cap\Omega_L}/\sizex{C}$ and $\rho_H$.

\begin{figure}[tb] 
\begin{center} 
\includegraphics[width=6cm,clip]{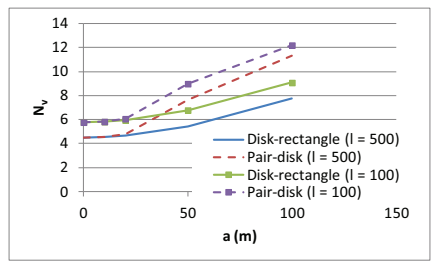} 
\caption{Impact of the shape of $D$ on $B_s$ and $N_v$} 
\label{shape} 
\end{center} 
\end{figure}

Finally, we investigate the impact of the shape of $D$ on $B_s(=B_d)$ and $N_v$ for fixed $\sizex{D}$.
(As a result, $a$ is variable while $r$ is given.)
Figure \ref{shape} shows the results.
For disk-rectangular $D$ and for pair-disk shaped $D$, $B_s$ is not very sensitive to the shape of $D$.
$N_v$ is more sensitive to the shape of $D$ than $B_s$.
This is because the length of $D$ as well as its size has a large impact on $N_v$, although $B_s$ is mainly determined by the size of $D$.
$N_v$ becomes minimum at $a=0$, that is, when $D$ is disk-shaped.
The difference in $N_v$ for disk-rectangles and for pair-disks becomes noticeable around $a\approx r$.
This is because $\lengthx{D}$ of a disk-rectangle $D$ becomes significantly different from that of a pair-disk $D$ when $a\gtrapprox r$.

\section{Conclusion}
We analyzed empirical data of the locations of WLAN APs and proposed an inhomogeneous Poisson process as a location model.
Based on the proposed model, explicit formulas for performance metrics, such as bandwidth (speed) available to users, were derived through integral geometry.
The derived formulas show good agreement with the simulation results using the empirical data.
They are exact under a homogeneous Poisson process.
We proved that the static available bandwidth and the dynamic available bandwidth are the same as a probabilistic distribution under a homogeneous Poisson process, and that their approximated probability are also the same when $\sizex{D_i}=\sizex{D}$ and $\lengthx{D_i}=\lengthx{D}$ for all $i$ even under an in homogeneous Poisson process.
These facts strongly suggest that the bandwidth experienced by a user is not dependent on user mobility.

Because these performance metrics depend on many parameters, such as the number of WLAN APs, the shape of each WLAN coverage region, the location of each WLAN AP, the available bandwidth (bps) of a WLAN, and the shape and available bandwidth (bps) of each subregion identified by the channel quality indicator in a cell of the cellular network, it is difficult to cover all the cases through computer simulation.
Therefore, the derived formulas are useful tools for performance evaluation of offloading through WLANs.

Numerical examples based on the derived formulas show the following:
(1) A homogeneous Poisson process can be too optimistic concerning the performance metrics such as user bandwidth (speed).
(2) Parameters, such as the size of regions where placement of WLAN APs is not allowed and the mean density of WLANs in high density regions, have a large impact on the performance metrics.

The analysis method used in this paper and some basic formulas derived in this paper are potentially applicable to other applications.
Actual applicability to other aplications will be investigated in the future.

\pagebreak
\appendix

\section{Proof of Eq. (\ref{g2})}\label{ap2}
Because $C_j \cap D_{i_1}\cap D_{i_2}\cap\cdots$ is convex, by applying Eq. (\ref{dDi}) for $i\neq i_1,\cdots,i_k$,
\begin{eqnarray}
&&\int_{\{\intersect{C}{D_i}\}_{i=1}^l}\sizex{C_j\cap_{m=i_1,\cdots,i_k}D_m }(dD)^l\cr
&\approx&\Pi_{i\neq i_1,\cdots,i_k} \rho_0g_1(C,D_i)\int_{\intersect{C}{D_{i_1}},\cdots,\intersect{C}{D_{i_k}}}\sizex{C_j\cap_{m=i_1,\cdots,i_k}D_m }dD_{i_1}\ \cdots\  dD_{i_k}\label{one}
\end{eqnarray}
Use Eq. (\ref{inhomo0}) with $\phi=\sizex{C_j\cap_{m=i_1,\cdots,i_k}D_m },X_c=\{\intersect{C}{D_i}\}_{i=1}^l$, and apply Eqs. (\ref{6.57-2}) and (\ref{6.57rr}). 
\begin{eqnarray}
&&\int_{\{\intersect{C}{D_i}\}_{i=1}^l}\sizex{C_j\cap_{m=i_1,\cdots,i_k}D_m }dD_{i_k}\cr
&=&\rho_0\{\int_{\{\intersect{C}{D_i}\}_{i=1}^l,\bfw_{i_k}\in {\mathbb R}^2}\sizex{C_j\cap_{m=i_1,\cdots,i_k}D_m }d\bfw_{i_k}\  d\gamma_{i_k}\cr
&&+\rho_H\int_{\{\intersect{C}{D_i}\}_{i=1}^l,\bfw_{i_k}\in \Omega_H}\sizex{C_j\cap_{m=i_1,\cdots,i_k}D_m }d\bfw_{i_k}\  d\gamma_{i_k}\cr
&&-\rho_L\int_{\{\intersect{C}{D_i}\}_{i=1}^l,\bfw_{i_k}\in \Omega_L}\sizex{C_j\cap_{m=i_1,\cdots,i_k}D_m }d\bfw_{i_k}\  d\gamma_{i_k}\}\cr
&\approx&2\pi\rho_0\sizex{D_{i_k}}(\sizex{C_j\cap_{m=i_1,\cdots,i_{k-1}}D_m }\cr
&&\qquad +\rho_H\sizex{C_j\cap_{m=i_1,\cdots,i_{k-1}}D_m\cap\Omega_H}-\rho_L\sizex{C_j\cap_{m=i_1,\cdots,i_{k-1}}D_m\cap\Omega_L}).
\end{eqnarray}
Again, due to Eqs. (\ref{6.57-2}), (\ref{6.57rr}), and (\ref{inhomo0}),
\begin{eqnarray}
&&\int_{\{\intersect{C}{D_i}\}_{i=1}^l}\sizex{C_j\cap_{m=i_1,\cdots,i_{k-1}}D_m\cap\Omega_H}dD_{i_{k-1}}\cr
&\approx&2\pi\rho_0\sizex{D_{i_{k-1}}}(\sizex{C_j\cap_{m=i_1,\cdots,i_{k-2}}D_m\cap\Omega_H}+\rho_H\sizex{C_j\cap_{m=i_1,\cdots,i_{k-2}}D_m\cap\Omega_H})\cr
&=&2\pi\rho_0(1+\rho_H)\sizex{D_{i_{k-1}}}\cdot\sizex{C_j\cap_{m=i_1,\cdots,i_{k-2}}D_m\cap\Omega_H},
\end{eqnarray}
because $\Omega_H\cap\Omega_H=\Omega_H$ and $\Omega_H\cap\Omega_L=\emptyset$.
Therefore,
\begin{eqnarray}
&&\int_{\{\intersect{C}{D_i}\}_{i=1}^l}\sizex{C_j\cap_{m=i_1,\cdots,i_{k-1}}D_m\cap\Omega_H}dD_{i_1}\cdots dD_{i_{k-1}}\cr
&\approx&\{2\pi\rho_0(1+\rho_H)\}^{k-1}\Pi_{m=1}^{k-1}\sizex{D_{i_m}}\cdot\sizex{C_j\cap\Omega_H}.
\end{eqnarray}
Similarly,
\begin{eqnarray}
&&\int_{\{\intersect{C}{D_i}\}_{i=1}^l}\sizex{C_j\cap_{m=i_1,\cdots,i_{k-1}}D_m\cap\Omega_L}dD_{i_1}\cdots dD_{i_{k-1}}\cr
&\approx&\{2\pi\rho_0(1-\rho_L)\}^{k-1}\Pi_{m=1}^{k-1}\sizex{D_{i_m}}\cdot\sizex{C_j\cap\Omega_L}.
\end{eqnarray}
Hence,
\begin{eqnarray}
&&\int_{\{\intersect{C}{D_i}\}_{i=1}^l}\sizex{C_j\cap_{m=i_1,\cdots,i_{k}}D_m }dD_{i_1}\ \cdots\  dD_{i_k}\cr
&\approx&2\pi\rho_0\sizex{D_{i_k}}\int_{\intersect{C}{D_{i_1}},\cdots,\intersect{C}{D_{i_{k-1}}}}\sizex{C_j\cap_{m=i_1,\cdots,i_{k-1}}D_m }dD_{i_1}\ \cdots\  dD_{i_{k-1}}\cr
&&\qquad +\rho_H(1+\rho_H)^{k-1}\Pi_{m=1}^{k}(2\pi\rho_0\sizex{D_{i_m}})\sizex{C_j\cap\Omega_H}-\rho_L(1-\rho_L)^{k-1}\Pi_{m=1}^{k}(2\pi\rho_0\sizex{D_{i_m}})\sizex{C_j\cap\Omega_L}.
\end{eqnarray}
Define 
\begin{eqnarray*}
J_k&\defeq&\int_{\intersect{C}{D_{i_1}},\cdots,\intersect{C}{D_{i_{k}}}}\sizex{C_j\cap_{m=i_1,\cdots,i_{k}}D_m }dD_{i_1}\ \cdots\  dD_{i_k}\cr
&&\qquad \qquad -(1+\rho_H)^{k}\Pi_{m=1}^{k}(2\pi\rho_0\sizex{D_{i_m}})\sizex{C_j\cap\Omega_H}\cr
&&\qquad \qquad -(1-\rho_L)^{k}\Pi_{m=1}^{k}(2\pi\rho_0\sizex{D_{i_m}})\sizex{C_j\cap\Omega_L}.
\end{eqnarray*}
Then, the equation above can be described as $J_k\approx 2\pi\rho_0\sizex{D_{i_k}}J_{k-1}$.
Therefore, $J_k\approx (2\pi\rho_0)^k\Pi_{j=1}^k \sizex{D_{i_j}}J_0=(2\pi\rho_0)^k\Pi_{j=1}^k \sizex{D_{i_j}}(\sizex{C_j}-\sizex{C_j\cap\Omega_H}-\sizex{C_j\cap\Omega_L})$.
\begin{eqnarray}
&&\int_{\{\intersect{C}{D_i}\}_{i=1}^l}\sizex{C_j\cap_{m=i_1,\cdots,i_{k}}D_m }dD_{i_1}\ \cdots\  dD_{i_k}\cr
&=&g_2(C_j,k)\Pi_{i= i_1,\cdots,i_k}(2\pi\rho_0\sizex{D_{i}})
\end{eqnarray}
By applying this result to Eq. (\ref{one}), we obtain Eq. (\ref{g2}).

\end{document}